# Phase-resolved Higgs response in superconducting cuprates

Hao Chu[1,2†], Min-Jae Kim[1,2], Kota Katsumi[3], Sergey Kovalev[4], Robert David Dawson[1], Lukas Schwarz[1], Naotaka Yoshikawa[3], Gideok Kim[1], Daniel Putzky[1], Zhi Zhong Li[5], Hélène Raffy[5], Semyon Germanskiy[4], Jan-Christoph Deinert[4], Nilesh Awari[4,6], Igor Ilyakov[4], Bertram Green[4], Min Chen[4,7], Mohammed Bawatna[4], Georg Christiani[1], Gennady Logvenov[1], Yann Gallais[8], Alexander V. Boris[1], Bernhard Keimer[1], Andreas Schnyder[1], Dirk Manske[1], Michael Gensch[7,9], Zhe Wang[4‡*], Ryo Shimano[3,10*], Stefan Kaiser[1,2*]

[1]*Max Planck Institute for Solid State Research, Heisenbergstr. 1, 70569 Stuttgart, Germany*

[2]*4th Physics Institute, University of Stuttgart, 70569 Stuttgart, Germany*

[3]*Department of Physics, University of Tokyo, Hongo, Tokyo,113-0033, Japan*

[4]*Helmholtz-Zentrum Dresden-Rossendorf, Bautzner Landstr. 400, 01328 Dresden, Germany*

[5]*Laboratoire de Physique des Solides (CNRS UMR 8502), Bâtiment 510, Université Paris-Saclay, 91405 Orsay, France*

[6]*University of Groningen, 9747 AG Groningen, Netherlands*

[7]*Technische Universität Berlin, Institut für Optik und Atomare Physik, Strasse des 17. Juni 135, 10623 Berlin, Germany*

[8]*Laboratoire Matériaux et Phénomènes Quantiques (UMR 7162 CNRS), Université de Paris, Bâtiment Condorcet, 75205 Paris Cedex 13, France*

[9]*German Aerospace Center (DLR), Institute of Optical Sensor Systems, Rutherfordstrasse 2, 12489 Berlin, Germany*

[10]*Cryogenic Research Center, University of Tokyo, Hongo, Tokyo,113-0032, Japan*

*email: zhewang@ph2.uni-koeln.de, shimano@phys.s.u-tokyo.ac.jp, s.kaiser@fkf.mpg.de

[†] Present address: Stewart Blusson Quantum Matter Institute, University of British Columbia, Vancouver, BC V6T 1Z4, Canada

[‡] Present address: Institute of Physics II, University of Cologne, 50937 Cologne, Germany

## Abstract

**In high energy physics, the Higgs field couples to gauge bosons and fermions and gives mass to their elementary excitations. Experimentally, such couplings can be inferred from the decay product of the Higgs boson, i.e. the scalar (amplitude) excitation of the Higgs field. In superconductors, Cooper pairs bear a close analogy to the Higgs field. Interaction between the Cooper pairs and other degrees of freedom provides dissipation channel for the amplitude mode, which may reveal important information about the microscopic pairing mechanism. To this end, we investigate the Higgs (amplitude) mode of several cuprate thin films using phase-resolved terahertz third harmonic generation (THG). In addition to the heavily damped Higgs mode itself, we observe a universal jump in the phase of the driven Higgs oscillation as well as a non-vanishing THG above $T_c$. These findings indicate coupling of the Higgs mode to other collective modes and potentially a nonzero pairing amplitude above $T_c$.**

## Introduction

For field theories with U(1) continuous symmetry and respecting Lorentz invariance or particle-hole symmetry, spontaneous symmetry breaking gives rise to an order parameter with two orthogonal collective modes: the Goldstone mode along the azimuthal direction, and the Higgs mode along the radial direction[1]. In superconductors, discussions about the Higgs mode and the Higgs mechanism precede those in high energy physics[2,3]. However, their significance to superconductivity has been more slowly appreciated. Experimental detection of the Higgs mode has also been hampered by its lack of electric and magnetic dipole moments in most superconductors. Recently, it was proposed that the Higgs mode may reveal the superconducting gap symmetry and multiplicity[4-6], unveil coupled collective modes[7], or explain aspects of light-induced superconductivity[8]. Novel methods for exciting and detecting the superconducting Higgs mode were also demonstrated using ultrafast terahertz techniques in the meantime[9,10]. Specifically, free oscillations of the Higgs mode, with its characteristic frequency of 2Δ, can be launched by an ultrashort terahertz pulse quenching the free energy of the order parameter[9]. Alternatively, it could also be periodically driven at $2\omega$ through nonlinear coupling between the electromagnetic vector potential $\mathbf{A}(\omega)$ and the superconducting condensate[10-12]. The resulting free (driven) Higgs oscillation manifests itself in terahertz transmissivity as an oscillation at 2Δ ($2\omega$). In the latter scenario, the 2ω oscillation of the condensate interacts with the driving field $\mathbf{A}(\omega)$, leading to sum frequency generation or third harmonic generation (THG), which is resonantly enhanced when $2\omega = 2\Delta(T)$[10,13].

While both free and driven Higgs oscillations have been demonstrated in *s*-wave superconductors, the Higgs mode of *d*-wave superconductors is more complex. The continuous variation of Δ between 0 and $\Delta^{max}$ along different directions of the Brillouin zone leads to strong dephasing of the mode. This is compounded by the existence of quasiparticle excitations at arbitrarily low energies, which provide rapid decay channels and significantly damp the mode[14]. Terahertz pump optical probe experiments on $Bi_2Sr_2CaCu_2O_{8+x}$ single crystals have provided the first experimental evidence for an isotropic Higgs response of *d*-wave superconductors in the form of an $|\mathbf{A}|^2$ response in the condensate's optical reflectivity to a monocycle terahertz pulse[15]. On the other hand, periodically driving the Higgs oscillation would provide useful phase information that may indicate resonance and coupling to other modes. Such an experiment requires a multicycle, carrier-envelope phase-stable terahertz source with a narrow bandwidth and high electric field strength, which is provided by the TELBE superradiant undulator source at HZDR[16]. Using this facility, we investigate the THG response of optimally-doped $La_{1.84}Sr_{0.16}CuO_4$ ($T_c$ = 45 K), $DyBa_2Cu_3O_{7-x}$ ($T_c$ = 90 K), $YBa_2Cu_3O_{7-x}$ ($T_c$ = 88 K), and overdoped $Bi_2Sr_2CaCu_2O_{8+x}$ ($T_c$ = 65 K) thin films (Supplementary Note 1). Our experiment is performed with 0.7 THz driving frequency, with an electric field up to ~ 50 kV $cm^{-1}$ (Supplementary Note 2). In all of these samples, we observe an increase in THG intensity ($I_{TH}$) below $T_c$ that is consistent with a heavily damped Higgs oscillation. In addition, in LSCO(OP45), DyBCO(OP90) and YBCO(OP88), a universal jump in the relative phase between the THG response and the linear drive ($\Phi_{TH}$) at $T < T_c$ is observed, signifying the coupling of the Higgs mode to another collective mode. We also observe a nonzero $I_{TH}$ persisting above $T_c$ in all of the samples, which may indicate preformed Cooper pairs above $T_c$.

## Results

**THG response of d-wave superconductors.** To illustrate the THG response of *d*-wave superconductors, first we show terahertz transmission through LSCO(OP45). As Figure 1(a)-(d) shows, while the residual fundamental harmonic (FH) dominates the terahertz transmission above $T_c$, a large amplitude of third harmonic (TH) becomes visible below $T_c$. Moreover, FH transmission ($I_{FH}$) monotonically decreases with decreasing temperature as shown in Fig. 1(e). In comparison, TH intensity ($I_{TH}$) exhibits a maximum below $T_c$. In the context of superconductors, the nonlinear Meissner effect[17,18], charge density fluctuations (CDF)[19], and Higgs oscillations have been previously reported or discussed to give rise to THG. Unlike what is observed in this study, the nonlinear Meissner effect manifests as a narrow peak around $T_c$ in the 3$^{rd}$ order nonlinear current. It is often discussed in terms of nonlinear Josephson current and might be probing the phase response of weakly connected superconducting islands[1]. On the other hand, studies on *s*-wave and *d*-wave superconductors have suggested an anisotropic response from CDF, while THG from the fully symmetric ($A_{1g}$) Higgs oscillation is expected to be isotropic[15,19-21]. To distinguish between CDF and the $A_{1g}$ Higgs response, we performed THG polarization dependence measurements. An isotropic response is found to dominate (Fig. 2(a)(c)). In addition, we performed terahertz pump optical probe (TPOP) measurements similar to [[15]]. An $|\mathbf{A}(\omega)|^2$ response to the $\mathbf{A}(\omega)$ driving field is also seen in the condensate's optical reflectivity (Fig. 2(b)). These results may not uniquely identify but are consistent with a driven Higgs response to the multicycle terahertz pulse and its role in THG below $T_c$. The THG polarization dependence indicates that there might be a finite contribution from other sources such as CDF or additional nonlinear mechanisms not yet discussed or experimentally evidenced. Therefore, a full understanding of the different sources of THG in addition to the driven Higgs oscillations requires further experimental and theoretical efforts. For instance, it will be interesting to investigate the systematic doping dependence of the anisotropic response in THG and TPOP experiments to quantify the relative contributions between different symmetry components[15]. Finally, to ensure that the Higgs oscillation stays in the perturbative excitation regime, we performed fluence dependence measurements. An excellent agreement with the expected $I_{\mathrm{TH}} \propto I_{\mathrm{FH}}^3$ dependence is observed (Fig. 1(f)) (Supplementary Note 6).

**Temperature dependence of THG.** To look for potential resonance of the Higgs oscillation at $2\omega = 2\Delta(T)$, we performed detailed temperature dependence measurements as shown in Fig. 3(a)-(c). In LSCO(OP45) we observe a peak in $I_{\mathrm{TH}}$ near $0.6T_c$ as well as a smaller peak around $0.9T_c$. DyBCO(OP90) exhibits a similar peak in $I_{\mathrm{TH}}$ near $0.6T_c$. In comparison, YBCO(OP88) exhibits a sharp peak in $I_{\mathrm{TH}}$ near $0.9T_c$ and a hump around $T_c$. In BSCCO(OD65), a continuously increasing $I_{\mathrm{TH}}$ with decreasing temperature is observed (Supplementary Note 7). A careful examination of the transmitted $I_{\mathrm{FH}}$ reveals that the main peak in $I_{\mathrm{TH}}$ originates from the competition between a growing nonlinear response of the Higgs oscillation and an increasing screening of the driving field as temperature decreases (Supplementary Note 3). Therefore, the main peak is not a resonance feature. In fact, for optimally-doped cuprates, we expect $\Delta(T{=}0) \gtrsim 20$ meV and a steep onset of $\Delta$ at $T_c$. Therefore, the $2\omega = 2\Delta(T)$ resonance, if present, is expected to be satisfied immediately below $T_c$ for $\omega = 0.7$ THz (~ 3 meV). Moreover, as the Higgs mode of *d*-wave systems is heavily damped, a resonance peak in $I_{\mathrm{TH}}$ is expected to be significantly broadened. This motivates us to investigate the phase of the driven Higgs oscillation, which is expected to exhibit a prominent change across resonance even in the presence of strong damping (Supplementary Note 8).

**Phase evolution of THG below $T_c$.** In Fig. 3(d)-(f), we extracted the relative phase ($\Phi_{\mathrm{TH}}$) of the TH with respect to the FH signal (Supplementary Note 4). Despite dissimilar features in $I_{\mathrm{TH}}$, all three samples exhibit a similar response in $\Phi_{\mathrm{TH}}$. In particular, an abrupt jump of nearly $\pi$ happens at a temperature $T_\pi < T_c$. In YBCO(88), $T_\pi$ is in the range where $2\omega = 2\Delta(T)$ is expected to be satisfied. However, such a sharp phase jump is again inconsistent with the resonance of a heavily damped collective mode. In LSCO(OP45) and DyBCO(OP90), $T_\pi$ is significantly lower than $T_c$. In light of these, we do not attribute the universal phase jump to the $2\omega = 2\Delta(T)$ Higgs resonance. A more striking evidence for such an interpretation comes from the direction of the phase jump. Since the low-temperature regime corresponds to driving below resonance ($2\omega < 2\Delta(T)$) and the high-temperature regime to driving above resonance ($2\omega > 2\Delta(T)$), a resonance-like phase jump should evolve positively with temperature whereas the observed $\Phi_{\mathrm{TH}}$ jumps negatively with temperature.

## Discussion

To obtain an intuitive understanding of the phase response, we look at a driven coupled harmonic oscillators model. Whereas isolated oscillators exhibit a maximum in their amplitude and a positive phase jump $\lesssim \pi$ across resonance[13] (Supplementary Note 8), the coupled oscillators system develops an anti-resonance in addition to resonances. This manifests as a minimum in the amplitude of the driven oscillator, simultaneous with a phase jump in the negative direction (Fig. 4(a)). To more closely model our experiment, we fix the driving frequency but allow the energetics of the oscillators to depend on temperature (Fig. 4(b)) (Supplementary Note 8). By choosing their resonance frequencies as $\Delta(T)$ and $\delta\Delta(T)$ ($\delta < 1$), where $\Delta(T) = \sqrt{n_s(T)}$ ($n_s$ is the experimentally measured superfluid density in LSCO(OP45)), the model recaptures the essential features of LSCO(OP45) (Fig. 3(a)(d)). In YBCO(OP88) a smaller dip in $I_{\mathrm{TH}}(T)$ is seen while in DyBCO(OP90) a kink in $I_{\mathrm{TH}}(T)$ is observed at the $T_\pi$. This could be due to the sharpness of the phase jump in these two samples, causing the dip in $I_{\mathrm{TH}}(T)$ to be very narrow in temperature and buried between the measured temperature points. In the framework of the coupled oscillators model, this suggests that the coupled mode is less damped in the bilayer systems compared to the single-layer LSCO(OP45). A classical Fresnel analysis of the thin film effects excludes the possibility of the phase jump coming from linear shifts in $\Phi_{\mathrm{FH}}(T)$ and $\Phi_{\mathrm{TH}}(T)$ (Supplementary Note 5).

While our classical toy model does not aim to explain the microscopic origin of the observed THG response, it allows us to learn about the dynamics and energetics of the coupled mode. It suggests that the coupled mode is underdamped and has an energy scale comparable to the Higgs mode. Moreover, the energy of this mode and/or its coupling to the Higgs mode depends on temperature. Within these constraints, potential candidates for the coupled collective mode include paramagnons, collective charge fluctuations of the incipient charge density wave (CDW) order, or phonons. In particular, coupling between the Higgs mode and CDW excitations has been observed in $NbSe_2$ and $2H$-$TaS_2$ superconductors[22,23]. In cuprates, charge order exhibits a similar energy scale as the superconducting gap[24]. Evidence for the incipient charge order in a wide range of the phase diagram has also been reported[25]. On the other hand, paramagnons are strongly renormalized in the superconducting state into a sharp underdamped resonance mode in a similar energy window as the superconducting gap [26]. They are prominent candidates for mediating Cooper pairing in high-$T_c$ superconductors. Last but not least, strong electron-phonon coupling may also contribute a phononic character to the superconducting order parameter[27], leading to new amplitude mode(s) in the Higgs response[7]. Other exotic but potential candidates include the Bardasis-Schrieffer mode[28,29] and the anisotropic $A_{2g}/B_{1g}$ Higgs mode[4,6] (Supplementary Note 9), which are also collective modes of the superconducting order parameter. Future doping and magnetic field dependence studies may shed light on the identity of the coupled mode.

While the identity of the coupled mode demands further scrutiny, our experiment also reveals a non-vanishing THG response above $T_c$ in all samples. In Fig. 4(c), we plot the temperature dependence of $(I_{TH}/I_{FH}^3)^{1/4}$, which is theoretically predicted to be $\propto \Delta$ away from resonance[13]. Indeed, $(I_{TH}/I_{FH}^3)^{1/4}$ exhibits an order parameter-like temperature dependence below $T_c$. Surprisingly, it remains nonzero up to $T > 1.5T_c$, similar to the temperature regime where superconducting fluctuations are observed in Nernst effect measurements[30]. This is more clearly illustrated by the transmitted TH waveform from LSCO(OP45) at $T = 50$ K $> T_c$ (Fig. 4(d)). Our observation may indicate preformed Cooper pairs in cuprates without global phase coherence, or the intense terahertz field might enforce phase coherence above $T_c$[12]. The pseudogap, and its various ordered phases including the CDW order, may also play a role in THG above $T_c$[31,32].

While the coupled oscillators model offers an intuitive explanation for the unexpected phase jump in THG in several families of cuprates, it is a classical toy model and calls for a full quantum mechanical treatment of the subject to provide deeper insight. In parallel, future magnetic field and doping dependence investigations may further help unveil the nature of the coupled mode and the non-vanishing THG response above $T_c$. Sweeping the driving frequency may also help distinguish the different scenarios of *d*-wave gap closing or filling as temperature approaches $T_c$. While these initiatives are underway, our technique may also serve as a probe for non-equilibrium superconductivity[33,34]. With so many interesting prospects, we see a bright future for phase-resolved spectroscopy of collective modes in superconductors and beyond[35-38].

## Methods

**Sample growth and characterization.** The LSCO(OP45) and DyBCO(OP90) samples were grown by molecular beam epitaxy (MBE), and the YBCO(OP88) sample was grown by pulsed laser deposition (PLD) at the Max Planck Institute for Solid State Research. The LSCO(OP45) sample is 80 nm-thick on a $LaSrAlO_4$ (LSAO) substrate. The DyBCO(OP90) sample is 70 nm-thick on a $(LaAlO_3)_{0.3}(Sr_2TaAlO_6)_{0.7}$ (LSAT) substrate. The YBCO(OP88) sample is 200 nm-thick on a $NdGaO_3$ (NGO) substrate. The BSCCO(OD65) sample was grown by sputtering technique at Laboratoire de Physique des Solides. The BSCCO(OD65) sample is 160 nm thick on a MgO substrate. As shown in Supplementary Note 1, $T_c$ is determined from mutual inductance measurement for LSCO(OP45), DyBCO(OP90) and YBCO(OP88). $T_c$ of BSCCO(OD65) is determined from the drop in magnetic moment from SQUID measurement under zero-field cooling. We define $T_c$ as the onset of the drop in mutual inductance and magnetic moment during cooling.

**THG experiment.** The majority of the data presented in this study are measured using the experimental setup shown in Supplementary Note 2. For fluence dependence measurements, we add an additional 1.93 THz bandpass filter (BPF) before Polarizer 3 (P3) to suppress the fundamental harmonic (FH). For temperature dependence of third harmonic (TH) in BSCCO(OD65), we also add an additional 1.9 THz BPF before P3. For electro-optical sampling we used a 2 mm ZnTe crystal and 100 fs gate pulse with 800 nm central wavelength. Accelerator-based THz pump and the laser gating pulse have a timing jitter characterized by a standard deviation of ~ 20 fs. Synchronization was achieved through pulse-resolved detection.

## Data Availability

The data that support the findings of this study are available from the first author and the corresponding authors upon reasonable request. For raw pre-binned data that allow statistical analysis, request should be sent to HZDR via S. Ko.

## Code Availability

The numerical code used to calculate the results for this work is available from H. C. and L. S. upon reasonable request.

## References

1. Pekker, D. & Varma, C. Amplitude/Higgs Modes in Condensed Matter Physics. *Annu. Rev. Condens. Matter Phys.* **6**, 269–297 (2015).
2. Higgs, P. W. Broken Symmetries and the Masses of Gauge Bosons. *Phys. Rev. Lett.* **13**, 508-509 (1964).
3. Anderson, P. W. Plasmons, Gauge Invariance, and Mass. *Phys. Rev.* **130**, 439-442 (1963).
4. Barlas, Y. & Varma, C. M. Amplitude or Higgs modes in d-wave superconductors. *Phys. Rev. B* **87**, 054503 (2013).
5. Krull, H. *et al.* Coupling of Higgs and Leggett modes in non-equilibrium superconductors. *Nat. Comm.* **7**, 11921 (2016).

6. Schwarz, L. *et al.* Classification and characterization of nonequilibrium Higgs modes in unconventional superconductors. *Nat. Comm.* **11**, 287 (2020).
7. Murakami, Y. *et al*. Multiple amplitude modes in strongly coupled phonon-mediated superconductors. *Phys. Rev. B* **93**, 094509 (2016).
8. Buzzi, M. *et al.* Higgs-mediated optical amplification in a non-equilibrium superconductor. Preprint at https://arxiv.org/abs/1908.10879 (2019).
9. Matsunaga, R. *et al.* Higgs Amplitude Mode in the BCS Superconductors. *Phys. Rev. Lett.* **111**, 057002 (2013).
10. Matsunaga, R. *et al.* Light-induced collective pseudospin precession resonating with Higgs mode in a superconductor. *Science* **345**, 1145-1149 (2014).
11. Anderson, P. W. Random-Phase Approximation in the Theory of Superconductivity. *Phys. Rev.* **112**, 1900–1916 (1958).
12. Yang, X. *et al.* Lightwave-driven gapless superconductivity and forbidden quantum beats by terahertz symmetry breaking. *Nature Photonics* **13**, 707-713 (2019).
13. Tsuji, N. & Aoki, H. Theory of Anderson pseudospin resonance with Higgs mode in superconductors. *Phys. Rev. B* **92**, 064508 (2015).
14. Peronaci, F. *et al.* Transient Dynamics of d -Wave Superconductors after a Sudden Excitation. *Phys. Rev. Lett.* **115**, 257001 (2015).
15. Katsumi, K. *et al.* Higgs Mode in the d-Wave Superconductor $Bi_2Sr_2CaCu_2O_{8+x}$ Driven by an Intense Terahertz Pulse. *Phys. Rev. Lett.* **120**, 117001 (2018).
16. Green, B. *et al.* High-Field High-Repetition-Rate Sources for the Coherent THz Control of Matter. *Sci. Rep.* **6**, 22256 (2016).
17. Lee, S.-C. *et al.* Doping-dependent nonlinear Meissner effect and spontaneous currents in high-$T_c$ superconductors. *Phys. Rev. B* **71**, 014507 (2005).

18. Mircea, D. *et al.* Phase-sensitive harmonic measurements of microwave nonlinearities in cuprate thin films. *Phys. Rev. B* **80**, 144505 (2009).

19. Cea, T. *et al.* Nonlinear optical effects and third-harmonic generation in superconductors Cooper pairs versus Higgs mode contribution. *Phys. Rev. B* **93**, 180507 (2016).

20. Matsunaga, R. *et al.* Polarization-resolved terahertz third-harmonic generation in a single-crystal superconductor NbN: Dominance of the Higgs mode beyond the BCS approximation. *Phys. Rev. B* **96**, 020505 (2017).

21. Schwarz, L. & Manske, D. Theory of driven Higgs oscillations and third-harmonic generation in unconventional superconductors. Preprint at https://arxiv.org/abs/2001.05662 (2020).

22. Littlewood, P. B. & Varma, C. M. Amplitude collective modes in superconductors and their coupling to charge-density-wave. *Phys. Rev. B* **26**, 4883-4893 (1982).

23. Grasset, R. *et al.* Pressure-Induced Collapse of the Charge Density Wave and Higgs Mode Visibility in $2H$–$TaS_2$. *Phys. Rev. Lett.* **122**, 127001 (2019).

24. Loret, B. *et al.* Intimate link between charge density wave, pseudogap and superconducting energy scales in cuprates. *Nat. Phys.* **15**, 771-775 (2019).

25. Arpaia, R. *et al.* Dynamical charge density fluctuations pervading the phase diagram of a Cu-based high-$T_c$ superconductor. *Science* **365**, 906-910 (2019).

26. Fong, H. F. *et al.* Polarized and unpolarized neutron-scattering study of the dynamical spin susceptibility of $YBa_2Cu_3O_7$. *Phys. Rev. B* **54**, 6708-6720 (1996).

27. Yang, X. *et al.* Ultrafast nonthermal terahertz electrodynamics and possible quantum energy transfer in the $Nb_3Sn$ superconductor. *Phys. Rev. B* **99**, 094504 (2019).

28. Bardasis, A. & Schrieffer, J. R. Excitons and Plasmons in Superconductors. *Phys. Rev.* **121**, 1050-1062 (1961).

29. Müller, M *et al.* Collective modes in pumped unconventional superconductors with competing ground states. *Phys. Rev. B* **100**, 140501 (2019).

30. Xu, Z. A. *et al.* Vortex-like excitations and the onset of superconducting phase fluctuation in underdoped $La_{2-x}Sr_xCuO_4$. *Nature* **406**, 486-488 (2000).

31. Pépin, C. *et al.* Fluctuations and Higgs mechanisms in Under-Doped Cuprates: a Review. Preprint at *https://arxiv.org/abs/1906.10146* (2019).

32. Cea, T. *et al.* Nonrelativistic Dynamics of the Amplitude (Higgs) Mode in Superconductors. *Phys. Rev. Lett.* **115**, 157002 (2015).

33. Fausti, D. *et al.* Light-Induced Superconductivity in a Stripe-Ordered Cuprate. *Science* **331**, 189-191 (2011).

34. Mitrano, M. *et al.* Possible light-induced superconductivity in $K_3C_{60}$ at high temperature. *Nature* **530**, 461-464 (2016).

35. Chu, H. *et al.* A charge density wave-like instability in a doped spin–orbit-assisted weak Mott insulator. *Nat. Mat.* **16**, 200-203 (2017).

36. Jain, A. *et al.* Higgs mode and its decay in a two-dimensional antiferromagnet. *Nat. Phys.* **13**, 633-637 (2017).

37. Werdehausen, D. *et al.* Coherent order parameter oscillations in the ground state of the excitonic insulator $Ta_2NiSe_5$. *Sci. Adv.* **4**, eaap8652 (2018).

38. Mansart, B. *et al.* Coupling of a high-energy excitation to superconducting quasiparticles in a cuprate from coherent charge fluctuation spectroscopy. *Proc. Natl. Acad. Sci. USA* **110**, 4539-4544 (2013).

## Acknowledgments

We thank G. Khaliullin, H. Boschker, M. Minola and H. Aoki for valuable discussions, and the ELBE team for the operation of the TELBE facility. Funding: S.Ka. acknowledges support by the Ministerium für Wissenschaft, Forschung und Kunst Baden-Württemberg through the Juniorprofessuren-Programm and a fellowship by the Daimler und Benz Stiftung. R.S. acknowledges partial support by JSPS KAKENHI Grant Nos.18H05324, 15H02102, and the Mitsubishi Foundation. N.A., I.I., S.Ko, M.G. acknowledge support from the European Union's Horizon 2020 research and innovation program under grant agreement No. 737038 (TRANSPIRE). S.Ka, H.C., M.-J.K. acknowledge support by the Max Planck-UBC-U Tokyo Center for Quantum Materials.

## Author contributions

S.Ka., R.S., Z.W., M.G., D.M., A.S. conceived the project. S.Ka., R.S., H.C., S.Ko., M.G. developed the experimental plan. S.Ko. built the setup at the TELBE facility. Data was taken and evaluated together with S.G., J.-C.D., N.A., I.I., B.G., M.C., M.B., M.G., Z.W., H.C., M.-J.K., K.K., R.D.D., N.Y., R.S., and S.Ka.. R.D.D., K.K., H.C., M.-J.K., A.V.B. conducted London penetration depth measurements. H.C., K.K., M.-J.K., R.D.D., N.Y., Z.W. performed data analysis and interpreted the results together with S.Ka., A.S., R.S., and B.K.. L.S., H.C. performed theoretical modeling. G.K., D.P., Z.Z.L., H.R., G.C., G.L., Y.G. provided and characterized samples. H.C. together with A.S. and S.Ka. wrote the manuscript with inputs from all authors.

## Competing interests

The authors declare that they have no competing interests.

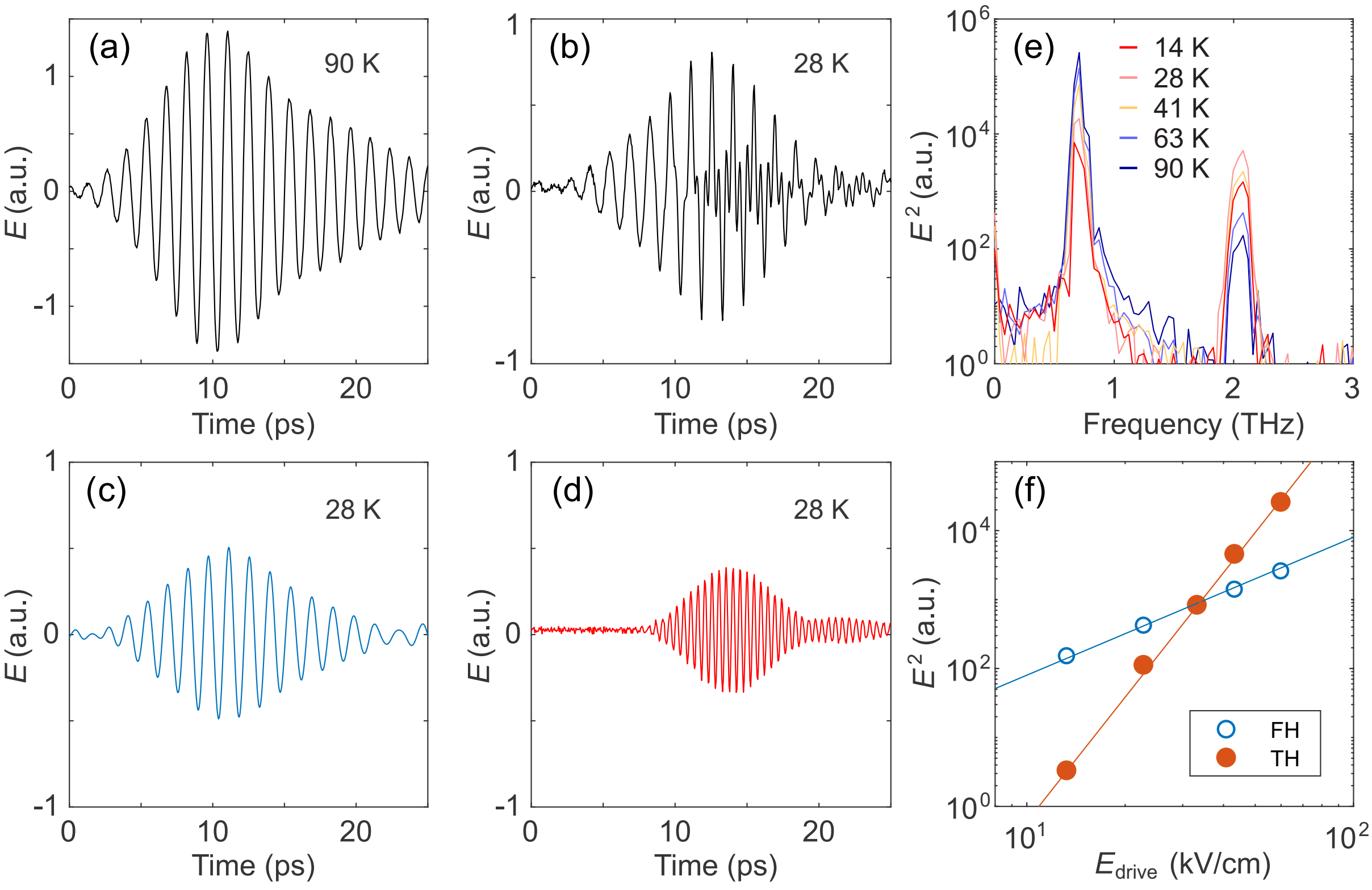

**Figure 1**

**THG from driven Higgs oscillation in LSCO(OP45). a,b** Terahertz field transmitted through LSCO(OP45) at 90 K and 28 K. A 2.1 THz bandpass filter is placed after the sample to suppress the 0.7 THz transmission (Supplementary Note 2). **c,d** 0.7 THz fundamental harmonic (FH) and 2.1 THz third harmonic (TH) extracted from **b** using 1.4 THz FFT low pass and high pass filters. **e,** FFT power spectrum of the transmitted field at selected temperatures across $T_c = 45$ K. **f,** Transmitted FH and TH power versus incoming FH field at 28 K. Solid lines are guides-to-the-eye with a slope of 2 and 6.

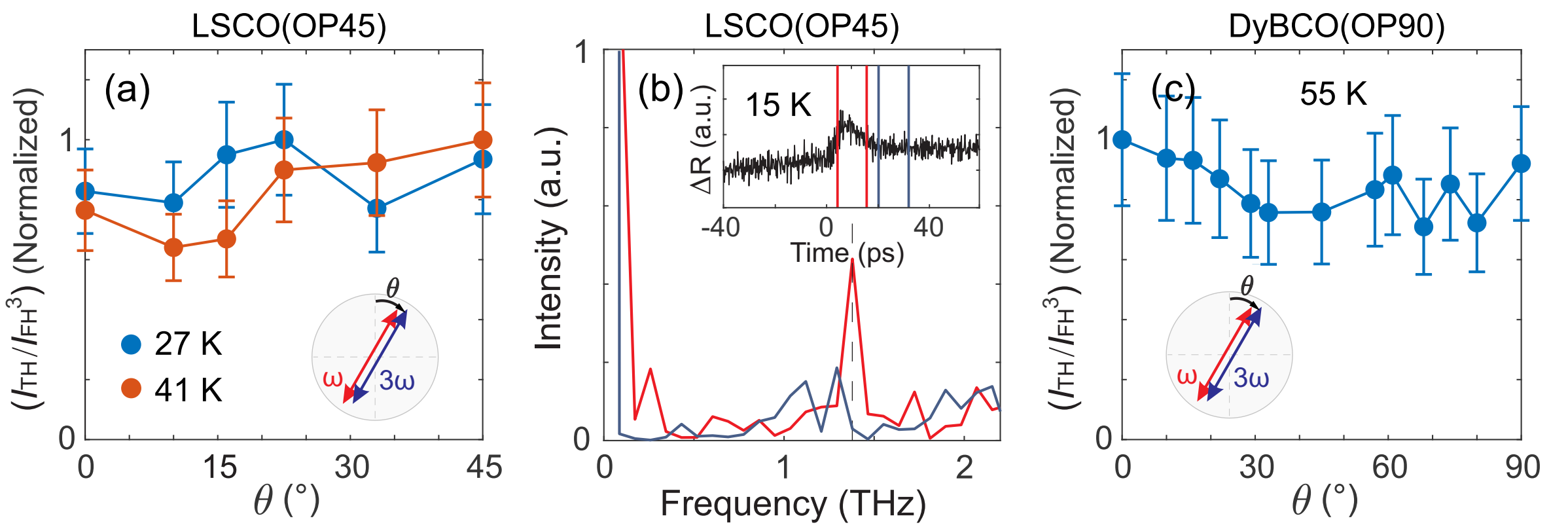

**Figure 2**

**Signature of the driven Higgs oscillations in transient response and polarization dependence. a,** THG intensity normalized by the parallel transmitted FH power ($I_{TH}/I_{FH}^3$) for LSCO(OP45) as a function of $\theta$, the angle from the Cu-O bond direction. THG from the Higgs oscillation is expected to be isotropic, while THG from charge density fluctuation (CDF) is expected to be anisotropic. **b,** transient reflectivity of LSCO(OP45) is measured with an 80 fs optical pulse while it is pumped with the 0.7 THz multicycle terahertz pulse. The change in reflectivity, ΔR, as a function of delay between the pump pulse and the probe pulse is shown in the inset. Main figure shows the FFT power spectrum of the relevant time intervals marked in the inset. ΔR exhibits a 1.4 THz (black dotted line) modulation while the pump pulse is on (red line) due to the $|\mathbf{A}|^2$ coupling of the condensate to the terahertz drive. The 1.4 THz peak becomes indistinguishable after the pump pulse is gone (blue line). **c,** THG polarization dependence for DyBCO(OP90). The error bars represent standard deviation due to pulse-to-pulse intensity fluctuation.

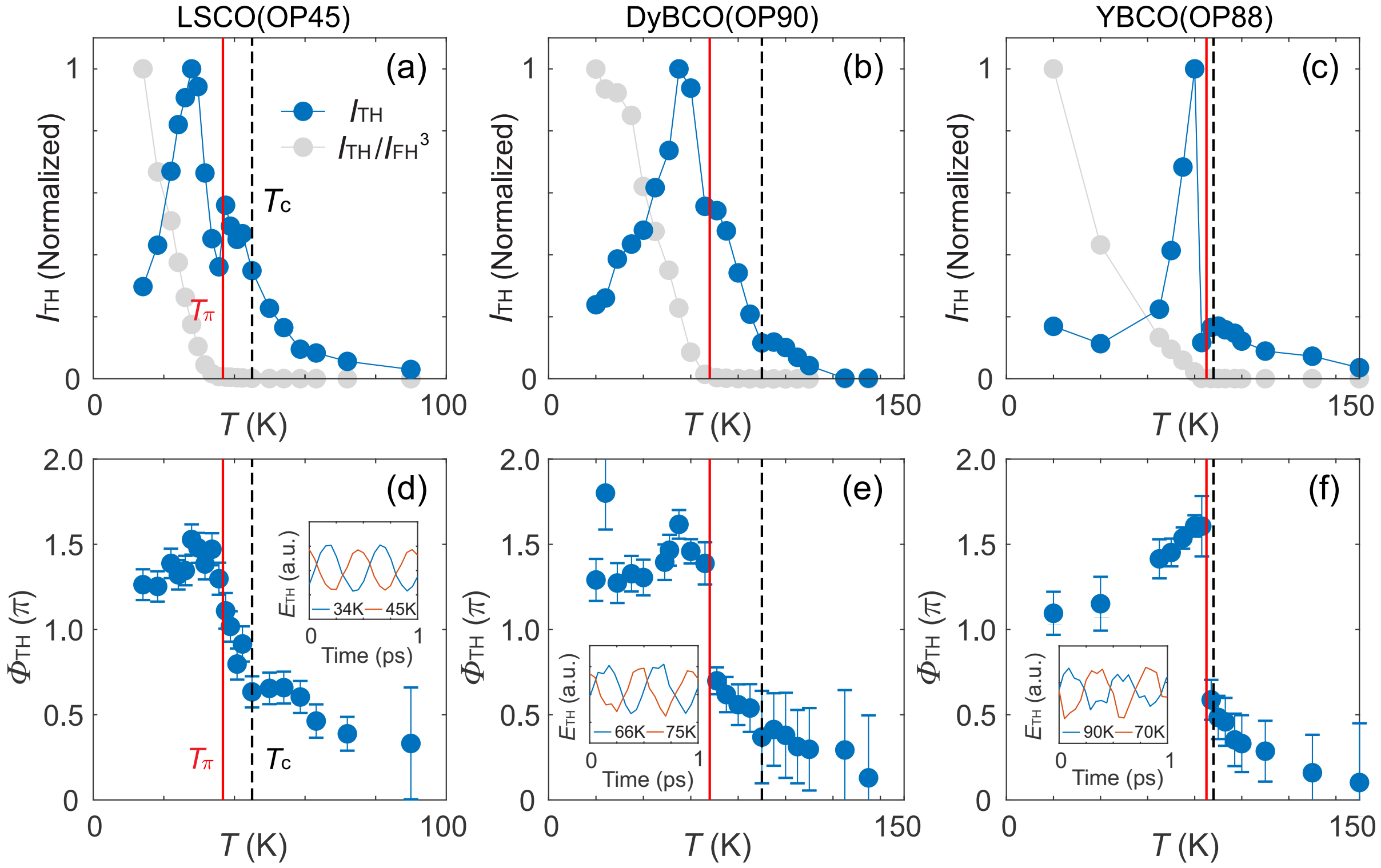

**Figure 3**

**Temperature dependence of TH intensity ($I_{TH}$) and relative phase ($\Phi_{TH}$) from optimally-doped cuprates. a-c,** Temperature dependence of TH intensity, $I_{TH}$ (blue), and normalized TH intensity, $I_{TH}/I_{FH}^3$ (grey), in LSCO(OP45), DyBCO(OP90) and YBCO(OP88). **d-f,** Temperature dependence of the relative phase between the TH response and the FH drive, extracted from waveforms such as those in Fig.1(c)(d). Inset shows representative TH waveforms across the π phase jump temperature ($T_\pi$). The dotted line (black) denotes $T_c$ and the solid line (red) denotes $T_\pi$. The error bars represent the two sigma uncertainty from fitting the phase of TH and FH waveforms (Supplementary Note 4).

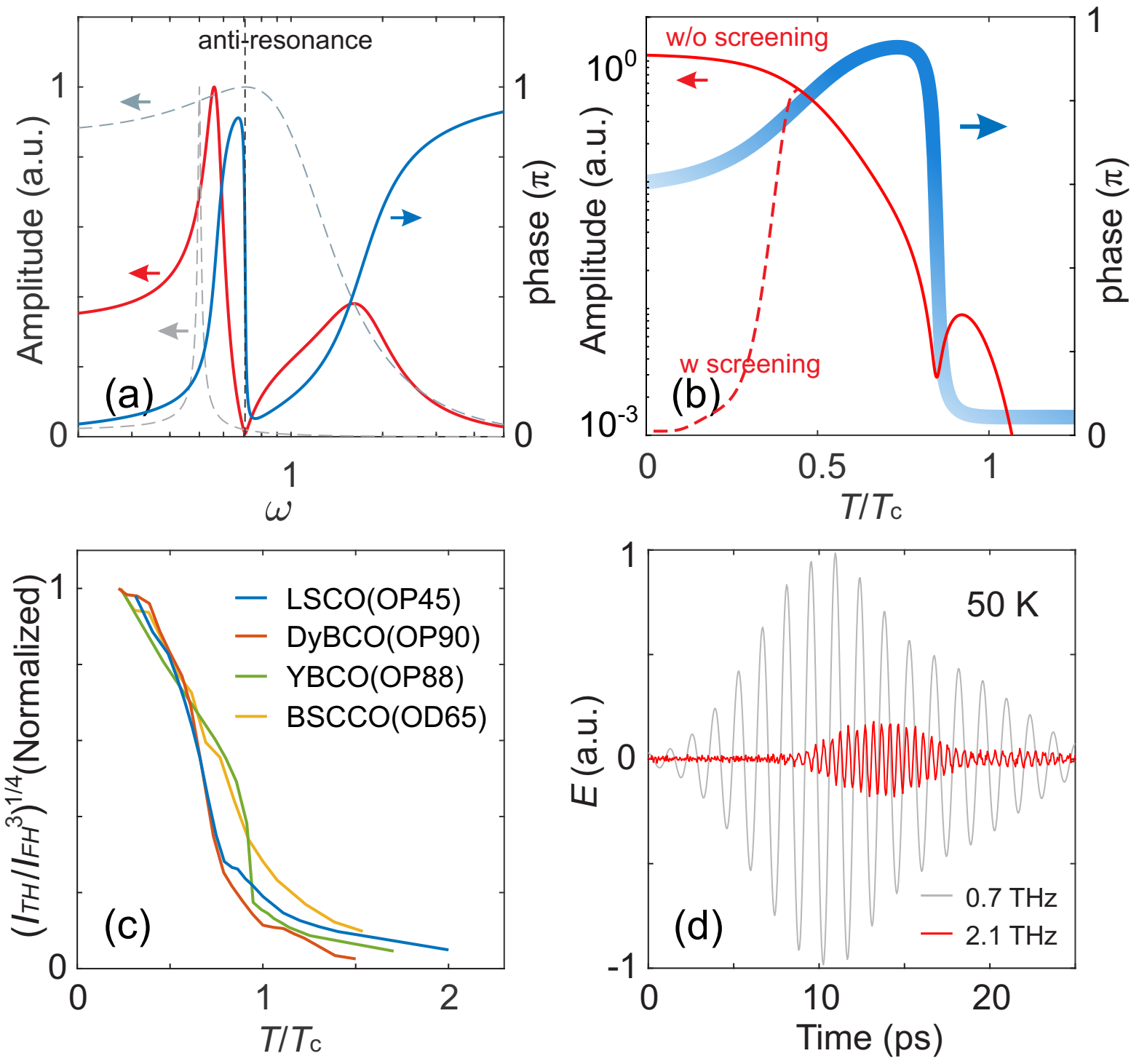

**Figure 4**

**Driven coupled oscillators model and finite THG above $T_c$. a,** the dotted lines depict the amplitude response of a critically damped harmonic oscillator and an underdamped harmonic oscillator as the driving frequency $\omega$ is varied. When these two oscillators are coupled, the coupled system retains two resonances but also develops an anti-resonance. Across the anti-resonance (vertical dotted line), the amplitude (solid red line) of the driven oscillator goes through a minimum while its phase (solid blue line) jumps negatively with the driving frequency. **b,** assuming the driving frequency is fixed as in our experiment, but that the resonance frequency of the two oscillators varies with temperature as $\Delta(T)$ and $\delta\Delta(T)$ ($\delta < 1$), the response of the coupled oscillators system is shown as a function of $T$. The anti-resonance in **a** is recaptured: the amplitude of the Higgs oscillation (thin red line) goes through a minimum while its phase (thick blue line) jumps negatively with $T$. Dotted line illustrates the effect of screening, which is to reduce the driving force and hence the amplitude at lower $T$. **c,** Temperature dependence of $(I_{TH}/I_{FH}{}^{3})^{1/4}$, which remains finite above $T_c$. $(I_{TH}/I_{FH}{}^{3})^{1/4}$ is theoretically predicted to be $\propto \Delta$ away from resonance. **d,** FH and TH components extracted from the transmitted waveform from LSCO(OP45) at 50 K > $T_c$.

# Supplementary Information

# Phase-resolved Higgs response in superconducting cuprates

Hao Chu[1,2], Min-Jae Kim[1,2], Kota Katsumi[3], Sergey Kovalev[4], Robert David Dawson[1], Lukas Schwarz[1], Naotaka Yoshikawa[3], Gideok Kim[1], Daniel Putzky[1], Zhi Zhong Li[5], Hélène Raffy[5], Semyon Germanskiy[4], Jan-Christoph Deinert[4], Nilesh Awari[4,6], Igor Ilyakov[4], Bertram Green[4], Min Chen[4,7], Mohammed Bawatna[4], Georg Christiani[1], Gennady Logvenov[1], Yann Gallais[8], Alexander V. Boris[1], Bernhard Keimer[1], Andreas Schnyder[1], Dirk Manske[1], Michael Gensch[7,9], Zhe Wang[4], Ryo Shimano[3,10], Stefan Kaiser[1,2]

*[1]Max Planck Institute for Solid State Research, Heisenbergstr. 1, 70569 Stuttgart, Germany*

*[2]4th Physics Institute, University of Stuttgart, 70569 Stuttgart, Germany*

*[3]Department of Physics, University of Tokyo, Hongo, Tokyo,113-0033, Japan*

*[4]Helmholtz-Zentrum Dresden-Rossendorf, Bautzner Landstr. 400, 01328 Dresden, Germany*

*[5]Laboratoire de Physique des Solides (CNRS UMR 8502), Bâtiment 510, Université Paris-Saclay, 91405 Orsay, France*

*[6]University of Groningen, 9747 AG Groningen, Netherlands*

*[7]Technische Universität Berlin, Institut für Optik und Atomare Physik, Strasse des 17. Juni 135, 10623 Berlin, Germany*

*[8]Laboratoire Matériaux et Phénomènes Quantiques (UMR 7162 CNRS), Université de Paris, Bâtiment Condorcet, 75205 Paris Cedex 13, France*

*[9]German Aerospace Center (DLR), Institute of Optical Sensor Systems, Rutherfordstrasse 2, 12489 Berlin, Germany*

*[10]Cryogenic Research Center, University of Tokyo, Hongo, Tokyo,113-0032, Japan*

Supplementary Note 1. Sample growth and characterization

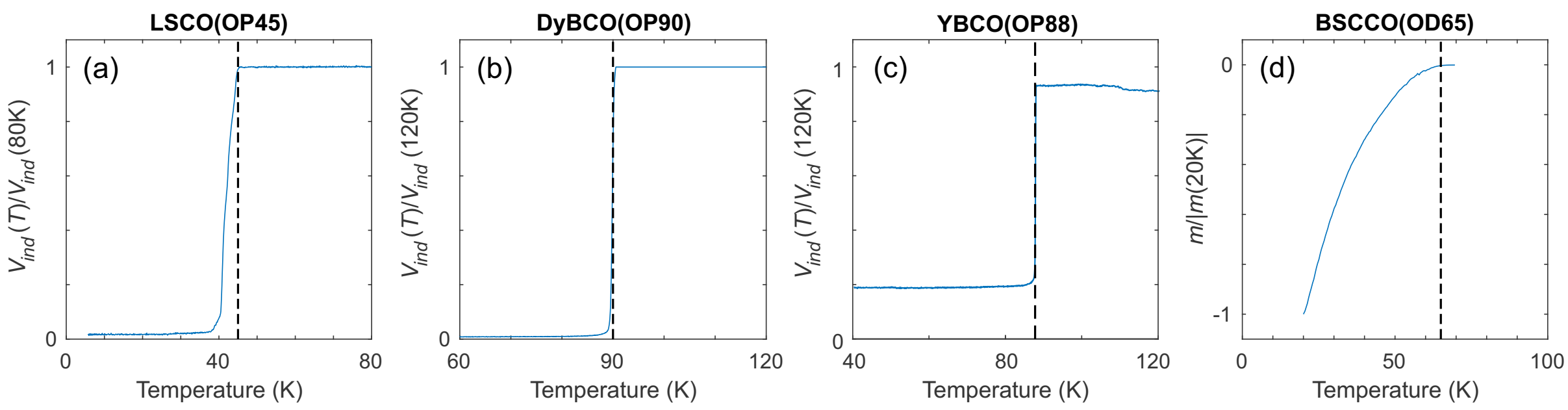

**Supplementary Figure 1** Experimental determination of $T_c$ in (a) LSCO (b) DyBCO and (c) YBCO by mutual inductance measurements and in (d) BSCCO by SQUID measurement. The mutual inductance results are normalized to their value above $T_c$. In BSCCO, the magnetic moment of the sample starts to drop at $T_c$. The drop is normalized to the magnetic moment at 20 K. Dotted line indicates $T_c$.

The LSCO(OP45) and DyBCO(OP90) samples were grown by molecular beam epitaxy (MBE), and the YBCO(OP88) sample was grown by pulsed laser deposition (PLD) at the Max Planck Institute for Solid State Research. The LSCO(OP45) sample is 80 nm-thick on a $LaSrAlO_4$ (LSAO) substrate. The DyBCO(OP90) sample is 70 nm-thick on a $(LaAlO_3)_{0.3}(Sr_2TaAlO_6)_{0.7}$ (LSAT) substrate. The YBCO(OP88) sample is 200 nm-thick on a $NdGaO_3$ (NGO) substrate. The BSCCO(OD65) sample was grown by sputtering technique at Laboratoire de Physique des Solides. The BSCCO(OD65) sample is 160 nm thick on a MgO substrate.

As shown in Supplementary Figure 1, $T_c$ is determined from mutual inductance measurement for LSCO(OP45), DyBCO(OP90) and YBCO(OP88). $T_c$ of BSCCO(OD65) is determined from the drop in magnetic moment from SQUID measurement under zero-field cooling. We define $T_c$ as the onset of the drop in mutual inductance and magnetic moment during cooling.

Supplementary Note 2. Experimental setup

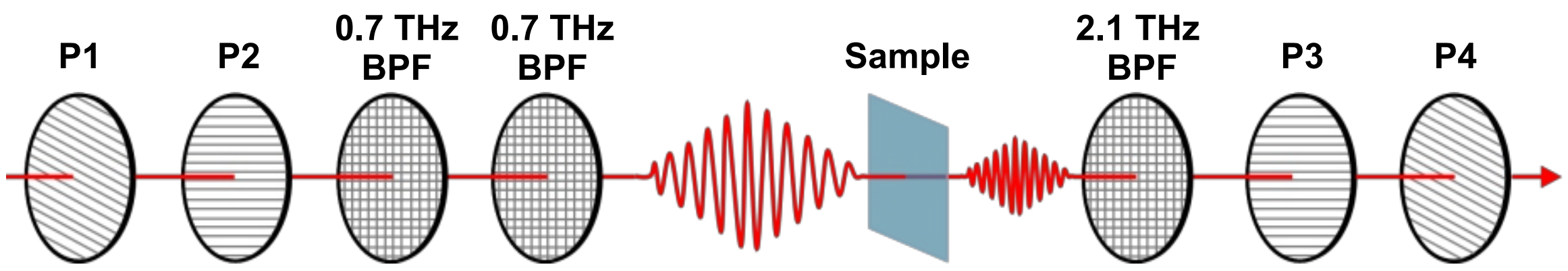

**Supplementary Figure 2.1** Experimental setup for the Higgs third harmonic generation experiment. P1-P4: wire-grid polarizers. BPF: bandpass filter.

The majority of the data presented in this study are measured using the experimental setup shown in Supplementary Figure 2.1. For fluence dependence measurements, we add an additional 1.93 THz BPF before P3 to suppress the fundamental harmonic (FH). For temperature dependence of third harmonic (TH) in BSCCO(OD65), we also add an additional 1.9 THz BPF before P3.

For electro-optical sampling we used a 2 mm ZnTe crystal and 100 fs gate pulse with 800 nm central wavelength. Accelerator-based THz pump and the laser gating pulse have a timing jitter characterized by a standard deviation of ~ 20 fs. Synchronization was achieved through pulse-resolved detection (*1*).

To estimate the efficiency of the third harmonic generation (THG), the 2.1 THz bandpass filter's transmission curve should be taken into account. The measured transmission of 2.1 THz BPF is shown in Supplementary Figure 2.2:

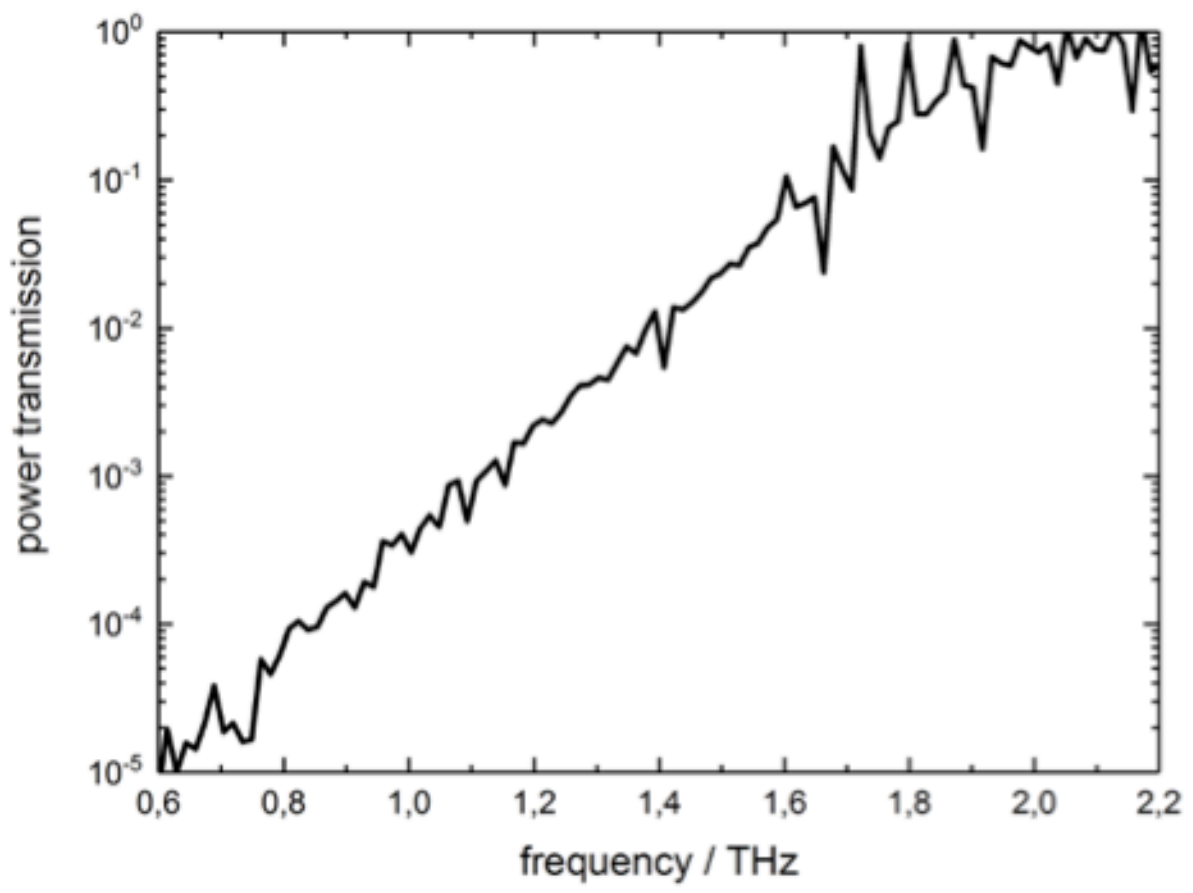

**Supplementary Figure 2.2** Power transmission of 2.1 THz bandpass filter

Supplementary Note 3. Temperature dependence of FH transmission

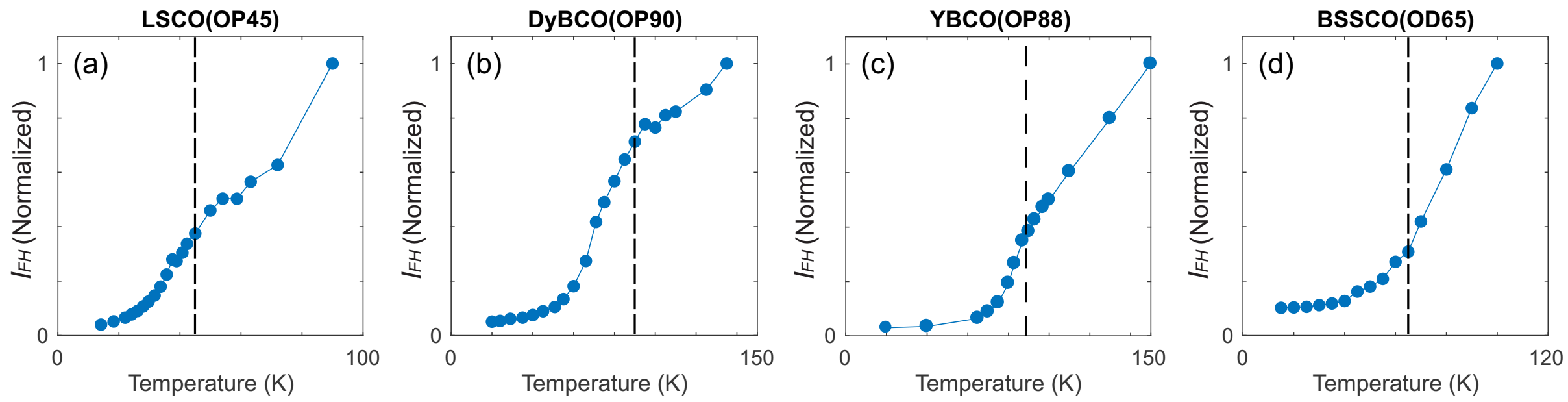

**Supplementary Figure 3.1** Temperature dependence of 0.7 THz FH transmission ($I_{\mathrm{FH}}$). The 0.7 THz transmission for (a) LSCO(OP45), (b) DyBCO(OP90), and (c) YBCO(OP88) are obtained from residual FH intensity in THG experiment. For (d) BSCCO(OD65), FH transmission is estimated from independent London penetration depth measurement.

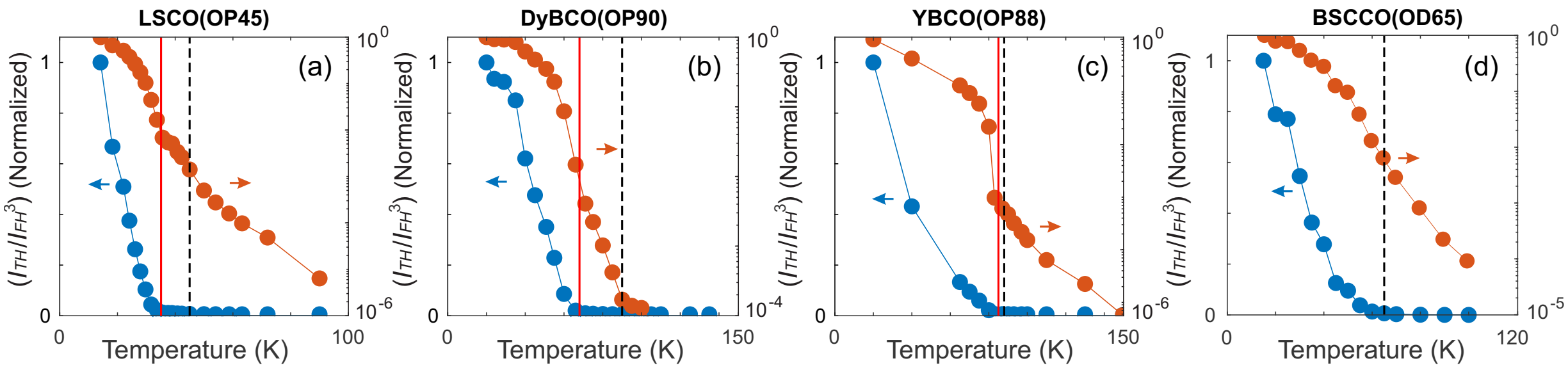

**Supplementary Figure 3.2** Temperature dependence of $I_{\mathrm{TH}}/I_{\mathrm{FH}}^3$ on linear scale (blue, left axis) and log scale (orange, right axis). Black dotted line denotes $T_c$, red solid line denotes $T_\pi$.

During the THG experiment, we measure the residual FH intensity ($I_{\mathrm{FH}}$) transmitted through the sample as a function of temperature. $I_{\mathrm{FH}}$ monotonically increases with temperature due to the decreasing screening effect of the superconducting condensate. For BSCCO(OD65), the transmitted FH intensity is very weak, resulting in high signal-to-noise ratio. Therefore, independent London penetration depth measurement is performed on BSCCO(OD65) to extract the transmission coefficient at 0.7 THz.

Taking the transmitted FH intensity as an estimate for the electric field inside the superconducting thin film and assuming $I_{\mathrm{TH}} \propto I_{\mathrm{FH}}^3$, we can correct for the screening effect and extract the intrinsic nonlinear response (susceptibility) of the Higgs mode. This is given by $I_{TH}/I_{FH}^3$ as shown in Supplementary Figure 3.2. The absence of a resonance-like peak near $T_c$, in contrast to THG from *s*-wave superconductors, is consistent with a heavily damped Higgs mode in *d*-wave superconductors. Note that the dip in $I_{TH}(T)$ at $T_\pi$ in LSCO(OP45) translates into a kink in $I_{TH}/I_{FH}^3(T)$. An even stronger kink is seen YBCO(OP88), while in DyBCO(OP90) it is less obvious.

Supplementary Note 4. Extraction of relative phase between TH response and FH drive

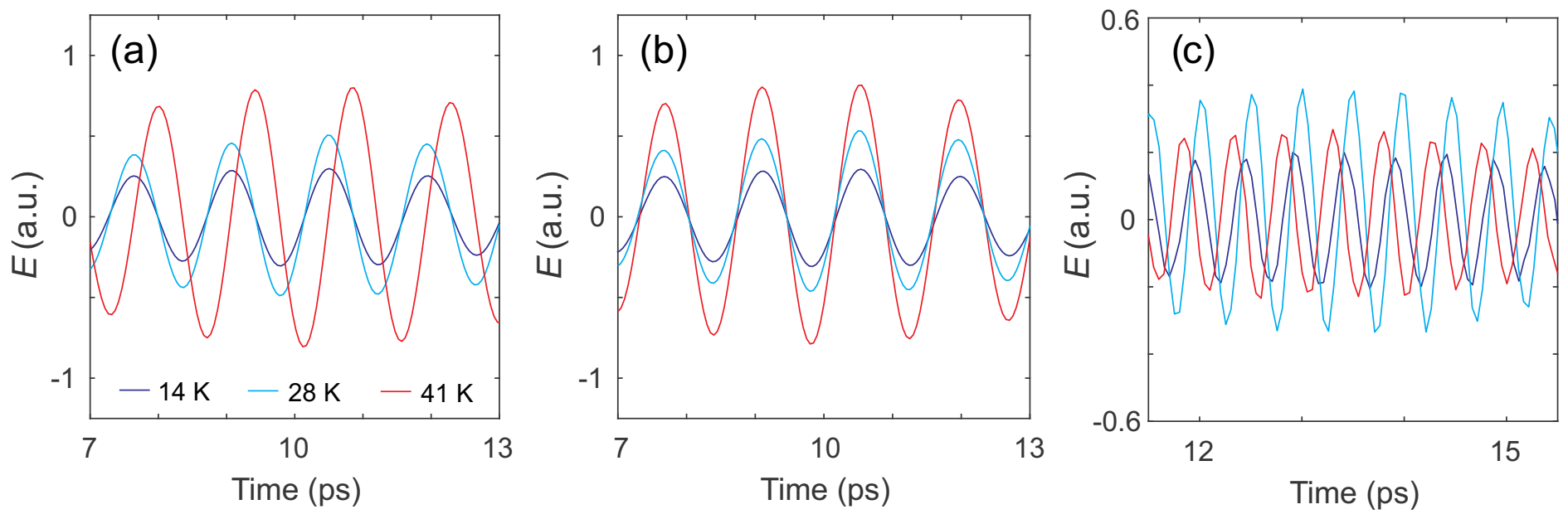

**Supplementary Figure 4** Procedure for extracting the relative phase between TH response and FH drive. **a,** raw FH waveforms at a few representative temperatures below $T_c$ in LSCO(OP45). **b,** FH waveforms aligned on top of each other after being shifted in time. **c,** TH waveforms at the same temperatures after applying the same time shift.

We first extract the raw FH and TH waveforms from the raw transmitted waveforms using 1.4 THz FFT low pass and high pass filters. For example, a few extracted FH waveforms from LSCO(OP45) are shown in Supplementary Figure 4(a). Due to the inductive response of superconductors below $T_c$, the FH wave experiences a phase shift across $T_c$ on transmission through the superconducting thin film. This is also illustrated in Supplementary Figure 4(a). As a first step, we apply a time shift $\delta t$ to the FH waveform at each temperature, so that their phases are all aligned with the lowest temperature waveform (Supplementary Figure 4(b)). Then, we apply the same time shift $\delta t$ to the corresponding TH waveform at each temperature. The resulting TH waveforms are shown in Supplementary Figure 4(c). Any phase shift between the TH waveforms in Supplementary Figure 4(c) has to intrinsically come from the Higgs oscillation itself because the phase shift in the FH drive has already been accounted for. To extracted this relative TH phase, we fitted the waveforms in Supplementary Figure 4(c) to a Gaussian-enveloped sinusoidal function,

$$\boldsymbol{E}_{\mathbf{TH}}(\boldsymbol{t}) = \boldsymbol{A}\,\mathbf{exp}\,(-(\boldsymbol{t}-\boldsymbol{t}_0)^2/\boldsymbol{c}^2)\,\mathbf{sin}(\boldsymbol{\omega}(\boldsymbol{t}-\boldsymbol{t}_0)-\boldsymbol{\Phi}),$$

where only $A$, $c$, $\boldsymbol{\Phi}$ are free fitting parameters. $\boldsymbol{\Phi}$ is the relative TH phase with respect to FH drive that we discussed extensively in the main text.

Supplementary Note 5. Effects of thin film on the intensity and phase of the transmitted fields

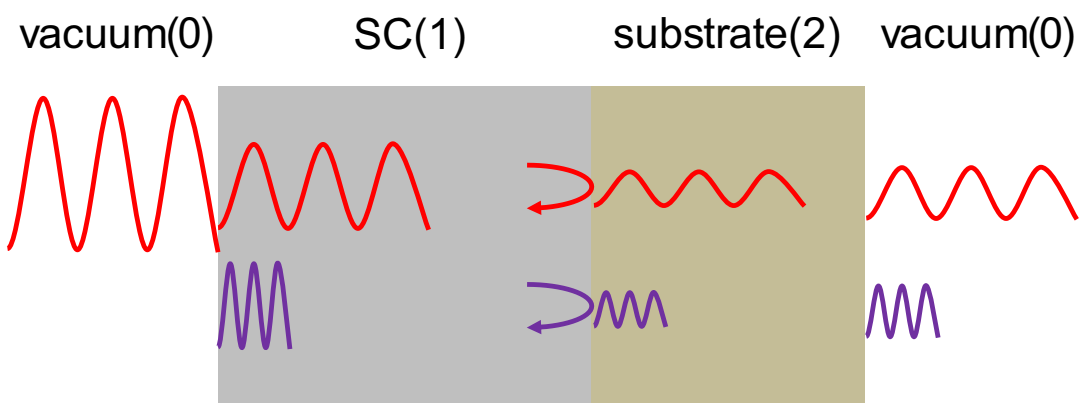

**Supplementary Figure 5.1** Illustration of electromagnetic wave propagation and THG in a thin film. Transmission and reflection at each interface is described by Fresnel equations and may depend on temperature and frequency. The thickness of the superconductor (SC) film is exaggerated for illustration purpose.

In our experiment, the FH field transmits through the sample composed of the superconductor (SC) film and the substrate, while TH field is presumably generated only within the SC film. The transmission of the FH wave through the entire sample is described by the Fresnel equations

$$E_{\text{measured}} = E_{\text{input}} t_{01} t_1 f_1 t_{12} t_2 t_{20} \qquad (1)$$

where $t_{\text{ij}} = \frac{2n_{\text{i}}}{n_{\text{i}}+n_{\text{j}}}$ describes transmission across interface from dielectric medium i to dielectric medium j, $t_{\text{i}} = \exp\,(i\omega L n_{\text{i}}/c)$ describes transmission through dielectric medium i of length $L$, $f_1 = \frac{1}{1+r_{01}r_{12}\exp\,(2i\omega L_1 n_1/c)}$ describes Fabry-Perot effect within the superconducting film of length $L_1$, $r_{\text{ij}} = \frac{n_{\text{i}}-n_{\text{j}}}{n_{\text{i}}+n_{\text{j}}}$ describes reflection at the interface from dielectric medium i to dielectric medium j. Here, $n_{\text{i}}$ denotes the complex index of refraction of medium i, $c$ denotes the speed of light in vacuum, and the subscript 0, 1, 2 refers to vacuum, SC, substrate respectively. Since $n_1$, the complex index of refraction of the SC film, strongly depends on temperature and frequency, transmission of the FH and TH field each develops its own temperature dependence. The measured FH and TH field outside the sample, therefore, may not be faithful representations of the FH and TH field inside the SC film. For example, the temperature dependence of $I_{\text{TH}}$, or $I_{\text{TH}}/I_{\text{FH}}^3$, using the value of $I_{\text{FH}}$ and $I_{\text{TH}}$ measured outside the sample may not give the real nonlinear response inside the SC film.

To account for such effect, we use the DyBCO(OP90) sample as an example and illustrate two different approaches. The first approach is based on back-calculating the electric field immediately on the left side of the SC-substrate interface from the electric field measured outside the sample. The second approach is based on simulating the propagation of the FH wave through the sample together with the THG process inside the SC film, and then predicting how the transmitted FH and TH field outside the sample would evolve as a function of temperature. Both methods show that the thin film has negligible effect on modifying the relative strength of the FH and TH field as well as the relative phase between them.

***First approach** (back-calculation)*

We are interested in knowing the exact FH and TH field inside the SC film. Since the thickness of the SC film is a very small fraction of the FH and TH wavelength and the FH and TH wave must be continuous inside the SC film, it is sufficient to know the FH and TH field, $E_1^{FH}$ and $E_1^{TH}$, immediately to the left of the SC-substrate interface. The measured electric field outside the sample is given by $E_{measured} = E_1\, t_{12}\, t_2\, t_{20}$ . Ignoring the temperature-independent factor ($t_2\, t_{20}$), we have $E_1^{FH} = E_{measured}^{FH}/t_{12}^{FH}$ and $E_1^{TH} = E_{measured}^{TH}/t_{12}^{TH}$, i.e. the electric field strength inside the SC film should be corrected from the electric field strength measured outside the sample by a factor $|t_{12}|$, and its phase from the measured phase by an offset $\arg(t_{12})$. Using the complex index of refraction of the DyBCO film and the LSAT substrate, we calculate $t_{12}^{FH}$ and $t_{12}^{TH}$. As shown in Supplementary Figure 5.2, the amplitude correction factor $|t_{12}|$ and the phase correction offset $\arg(t_{12})$ are both small and varies little with temperature.

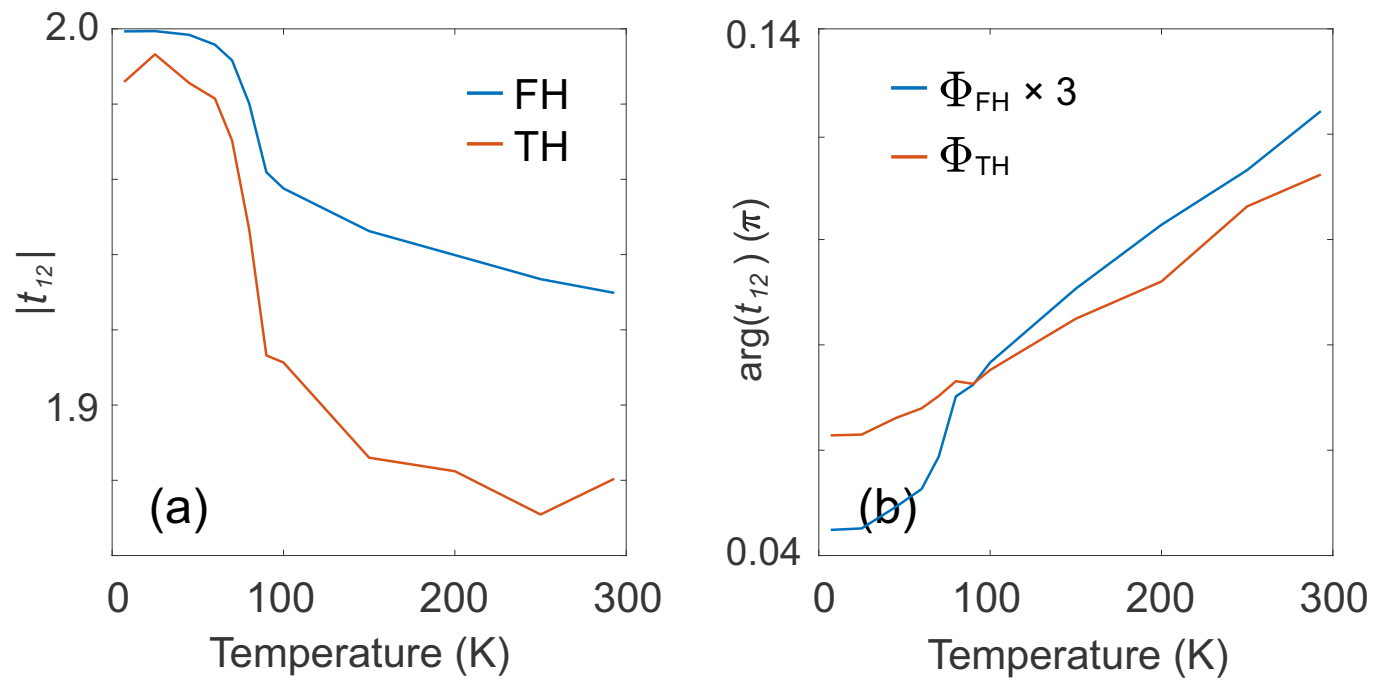

**Supplementary Figure 5.2** Amplitude and phase correction for the electric field measured outside the sample due to transmission across the SC-substrate interface.

***Second approach** (forward simulation)*

In this approach, we simulate the transmission of the FH wave through the entire sample and the THG process within the SC film. Below $T_c$, the vacuum-SC interface and the SC-substrate interface forms two highly reflective surfaces for terahertz frequencies. Therefore, the Fabry-Perot effect needs to be considered as well. This is illustrated in Supplementary Figure 5.3. The resulting FH field inside the SC film is the superposition of all the reflected FH waves. We then assume $E_1^{TH}$ is generated from this total FH field using the simple relation $E_1^{TH} = (E_1^{FH})^3$. The generated $E_1^{TH}$ also undergoes multiple reflections at the two interfaces. Therefore the total TH field inside the SC film is also the superposition of all the reflections. We then calculate the field transmitted outside the sample using $E_{measured} = E_1 t_{12}$, where $E_1$ is the electric field immediately to the left of the SC-substrate interface. The resulting time and temperature dependence of $E_{measured}^{FH}$ and $E_{measured}^{TH}$ is shown in Supplementary Figure 5.4. It can be seen that the linear effects arising from interfaces, the SC film, and Fabry-Perot effect cause the FH and TH waves to shift a similar amount in time. This is more clearly shown in Supplementary Figure 5.5(b). The Fabry-Perot effect also leads to small correction of the electric field strength inside the SC film. Ideally, since we assumed $E_1^{TH} = (E_1^{FH})^3$ for the THG process, we should have $|E_1^{TH}/(E_1^{FH})^3| = 1$. However, $|E_1^{TH}/(E_1^{FH})^3|$ deviates from 1 due to the Fabry-Perot effect for FH and TH waves respectively. This leads to a correction factor for the FH and TH field strength inside the SC film, which, however, also turns out to be very small: Supplementary Figure 5.5(a).

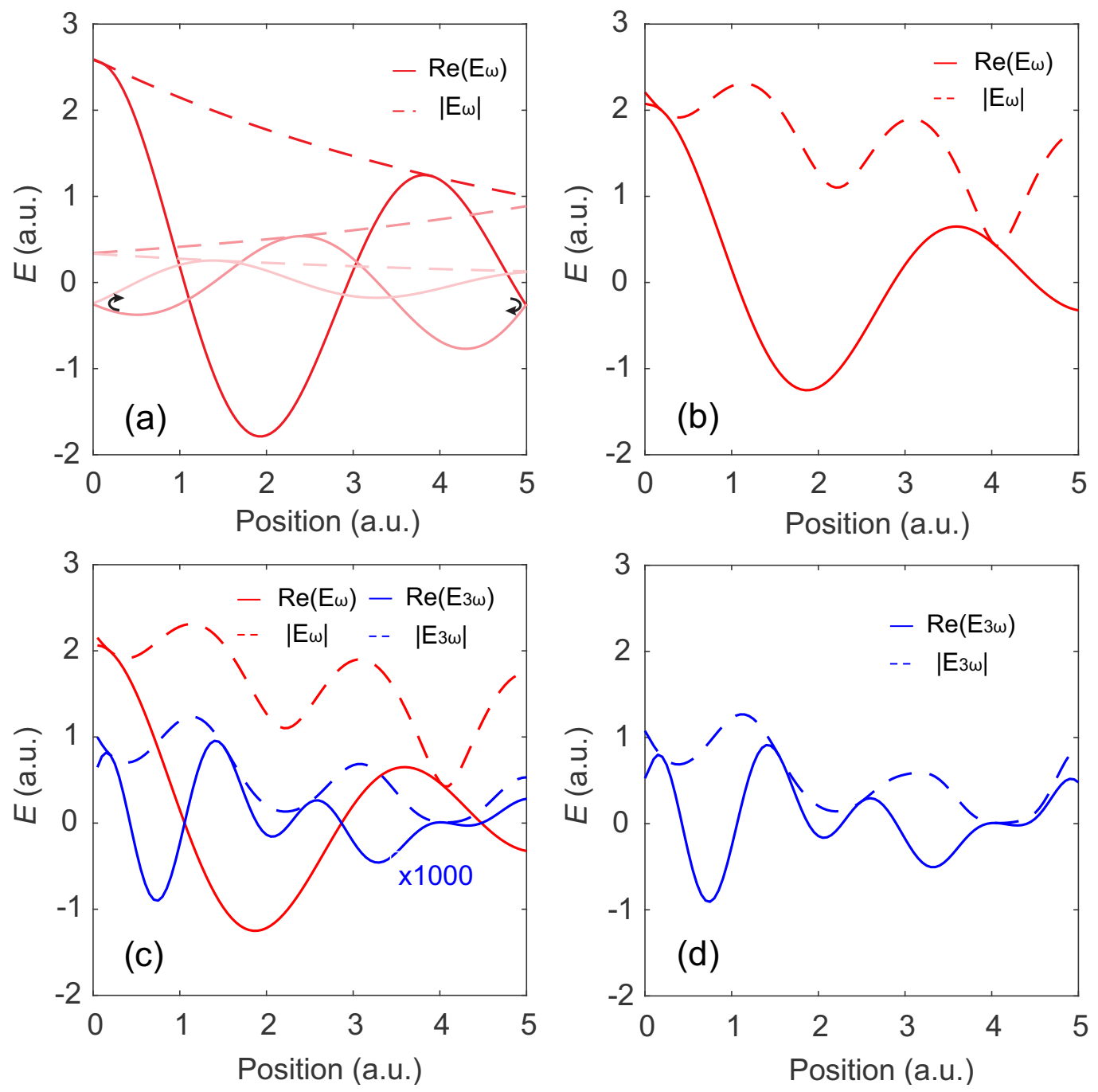

**<u>Supplementary Figure 5.3</u>** Illustration of FH transmission through the SC film and THG inside the SC film**. a,** multiple reflections of the FH wave inside the SC film. **b,** the resulting FH field which is the superposition of all reflected FH waves. **c,** THG from FH assuming $E_1^{\mathrm{TH}} = (E_1^{\mathrm{FH}})^3$. **d,** the resulting TH field which is the superposition of all reflected TH waves. For the purpose of illustration, the thickness of the SC film is exaggerated so as to accommodate 1.25 cycles of the FH wave. The actual SC film has a thickness on the order of $10^{-3}$ of the wavelength of the FH wave. The index of refraction is also purposely selected to illustrate absorption effect.

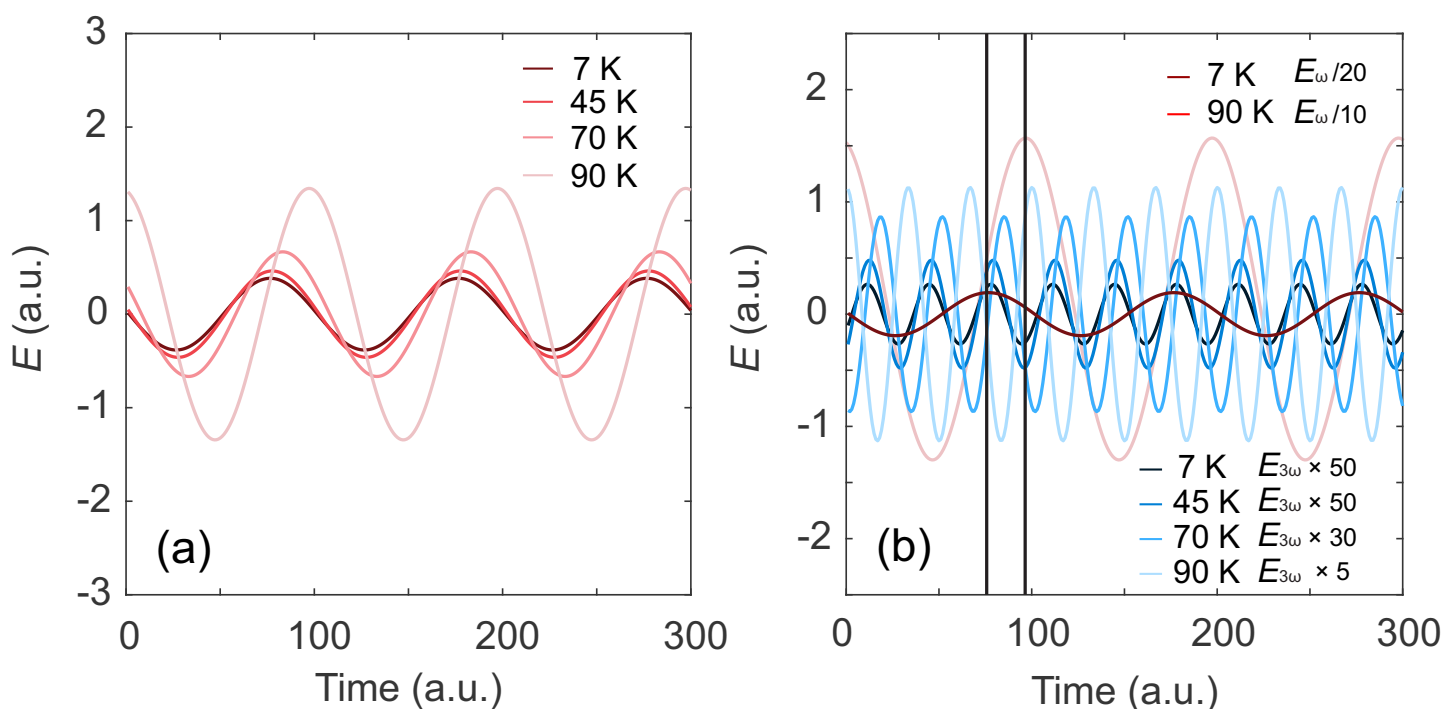

**Supplementary Figure 5.4** Transmitted FH and TH fields outside the sample as a function of time. **a,** the transmitted FH field, $E_{\mathrm{measured}}{}^{\mathrm{FH}}$, for several different temperatures. **b,** the transmitted TH field, $E_{\mathrm{measured}}{}^{\mathrm{TH}}$, for the same temperatures. The transmitted FH fields at 7 K and 90 K are superposed for comparison. Both FH and TH fields undergo a similar shift in time (designated by the two solid black lines) between 7 K and 90 K.

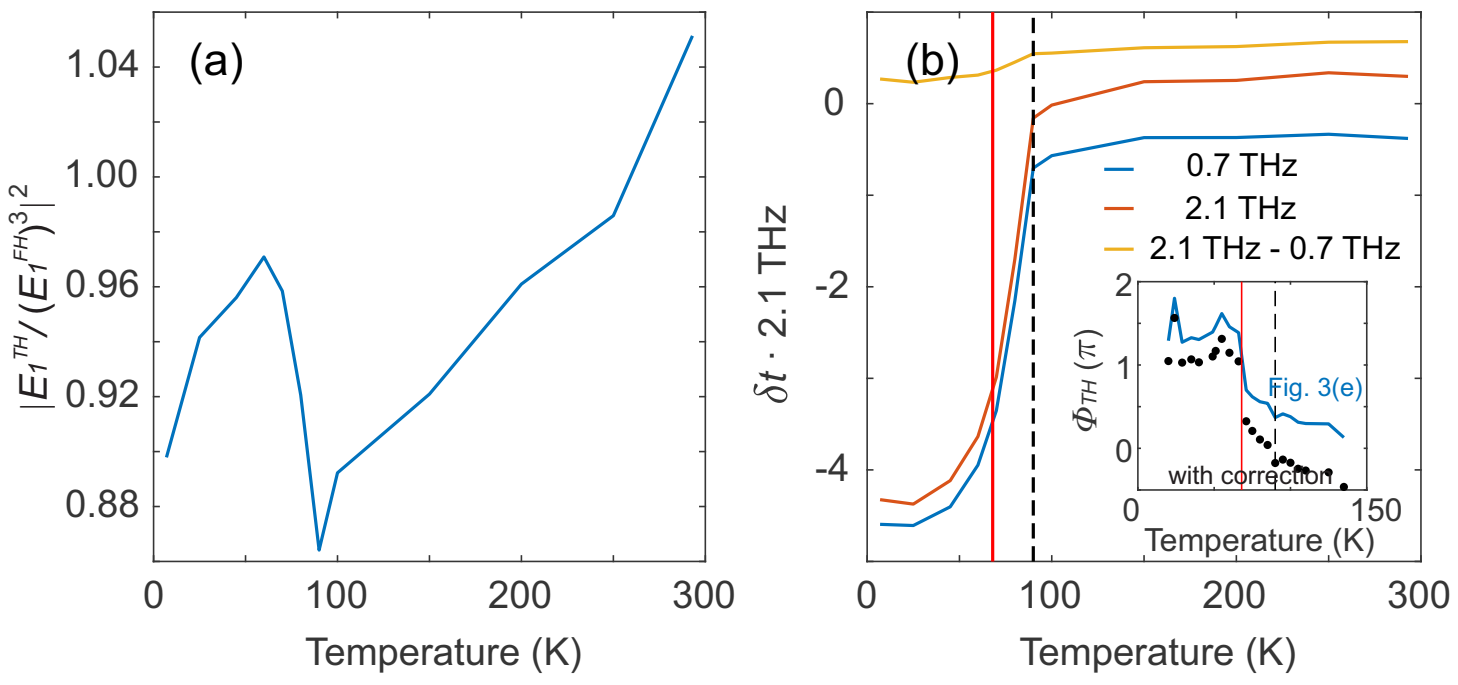

**Supplementary Figure 5.5** **a,** Correction factor for $I_{\mathrm{TH}}/I_{\mathrm{FH}}{}^3$ due to the Fabry-Perot effect. **b,** time-shift predicted for the transmitted FH and TH waves due to temperature dependent variation of linear optical properties. It can be seen that FH and TH field undergoes a similar shift in time. Therefore, the correction for the relative TH phase coming from thin film effects (yellow curve) is quite small. In the inset, we apply this correction to the relative TH phase in DyBCO(OP90). The corrected relative TH phase (black dots) retains the sharp $\pi$ jump near $T_\pi$ (red solid line) similar to the original relative TH phase from the main text (blue line). Black dotted line denotes $T_c$.

Supplementary Note 6. Fluence dependence of THG

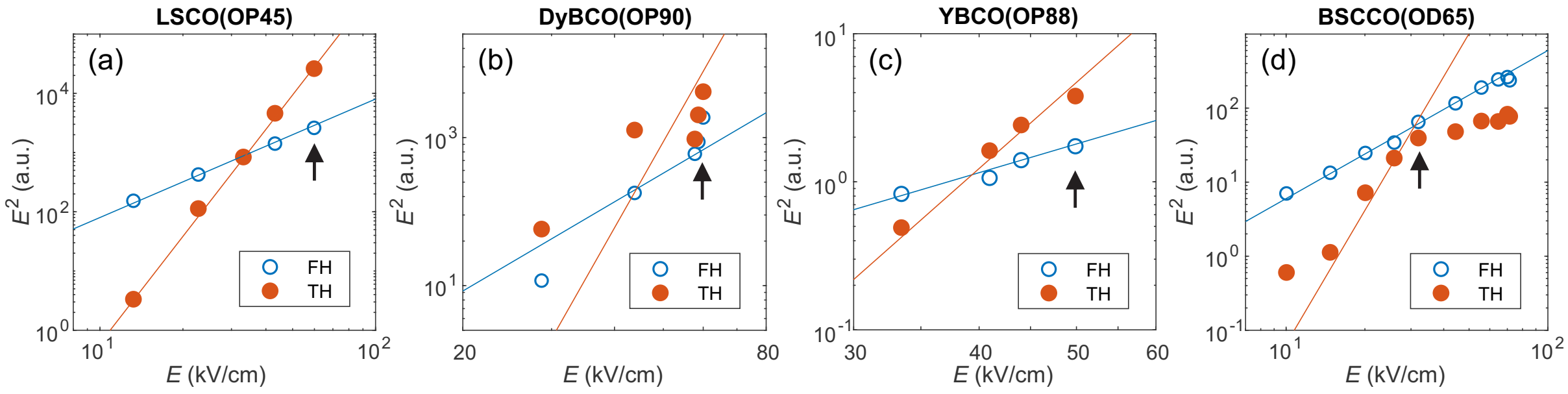

**Supplementary Figure 6** Fluence dependence of THG. The transmitted TH and FH intensity is measured as a function of the incoming FH field for (a) LSCO(OP45) at 27 K, (b) DyBCO(OP90) at 70 K, (c) YBCO(OP88) at 52 K, and (d) BSCCO(OD65) at 20 K. Solid lines are guides to the eye with a slope of 2 and 6. Arrows indicate the FH field with which the data in the main text are taken.

To make sure that the THG experiment stays within perturbative regime of the Higgs mode, we performed fluence dependence measurements. As shown in Supplementary Figure 6, LSCO(OP45) exhibits excellent agreement with the $I_{TH} \propto I_{FH}^3$ dependency. YBCO(OP88) has similar $I_{TH} \propto I_{FH}^3$ dependency. BSCCO(OD65) exhibits both a $I_{TH} \propto I_{FH}^3$ regime and non-$I_{TH} \propto I_{FH}^3$ regimes. In DyBCO(OP90), deviation from the $I_{TH} \propto I_{FH}^3$ dependency is the most pronounced. Based on the results of these measurements, we pick for our experiment the highest FH field within the perturbative regime to maximize signal-to-noise ratio.

Supplementary Note 7. THG in BSCCO(OD65)

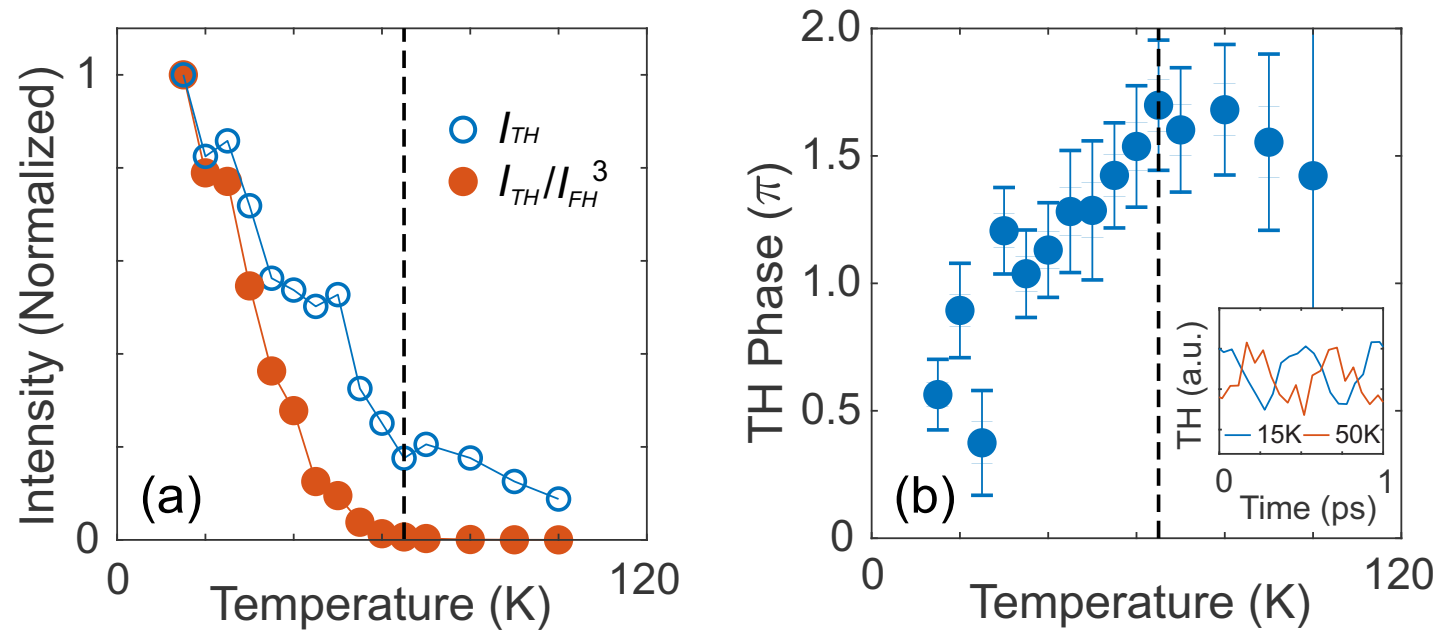

**Supplementary Figure 7** Temperature dependence of TH intensity and relative phase in BSCCO(OD65).

THG from BSCCO(OD65) is very weak, which can be inferred from the noisy TH waveforms extracted in Supplementary Figure 7(b). The temperature dependence of TH intensity shows a monotonically increasing trend towards low temperature. In addition, the relative TH phase shows a gradual π shift in a direction opposite to the π phase shift in optimally doped samples. These dramatically different results compared to the optimally doped samples might be due to overdoping, but it could also be that BSCCO(OD65) exhibits a more gradual superconducting transition due to sample inhomogeneity. The inhomogeneity likely stems from an oxygen gradient that has formed in the film during the time between the sample growth and the THG measurements. For these reasons, we refrain from making interpretations about the BSCCO(OD65) THG results.

Supplementary Note 8. Driven coupled harmonic oscillators model

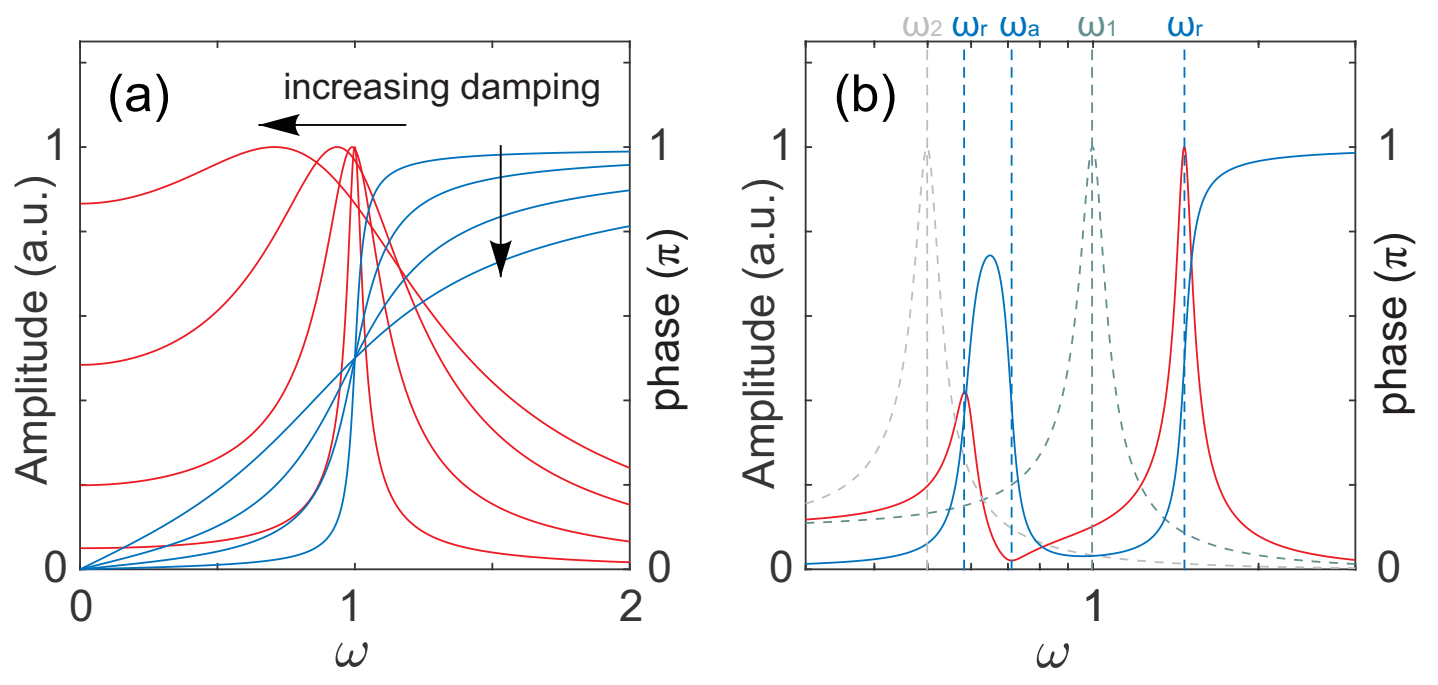

**Supplementary Figure 8.1** **a,** the response of an isolated driven harmonic oscillator as a function of driving frequency. As damping increases, the peak in amplitude (red lines) moves below the resonance frequency ($\omega_1 = 1$) while the center of the phase shift (blue lines) still intersects at the resonance frequency. **b,** dotted lines show the amplitude response of two isolated underdamped driven harmonic oscillators with resonance frequency $\omega_1$ and $\omega_2$. Solid line show the amplitude (red) and phase (blue) response when the two oscillators are coupled. $\omega_r$ denotes the resonance frequency of the coupled system, $\omega_a$ denotes the anti-resonance frequency of the coupled system. The parameters used for generating **b** are $\omega_1 = 1$, $\omega_2 = 0.5$, $b_1 = 0.1$, $b_2 = 0.05$, $\omega_{c1} = 1$, $\omega_{c2} = 0.5$.

We consider a model of two driven damped harmonic oscillators, each modeled by a spring constant $k_i$, a mass $m_i$, a damping coefficient $b_i$, and a spring constant $K$ that couples the two oscillators. Furthermore, we assume $\omega_1 = (k_1/m_1)^{1/2} = 2\Delta$, and $\omega_2 = (k_2/m_2)^{1/2} = \delta \times 2\Delta$, where $0 < \delta < 1$. Oscillator 1 represents the 2Δ Higgs mode, while oscillator 2 represents the other collective mode. Since light couples to the Higgs mode quadratically, the frequency of the periodic driving force, as felt by the Higgs mode, is 2ω. For simplicity, we drop off the factor of 2 from the driving force frequency (2ω) and the energies of the two oscillators (2Δ and $\delta \times 2\Delta$) in the following discussion. This way, the resonance condition for the Higgs mode at $2\omega = 2\Delta(T)$ is still preserved and given by $\omega = \Delta(T)$.

In the case that the two oscillators are decoupled, $K = 0$. The linear response of each oscillator is given by

$$x_i = \frac{\omega_i^2 X_i}{\omega_i^2 - \omega^2 + i b_i \omega}$$

where $X_i$ is the amplitude of the drive, and ω is the driving frequency. A noticeable effect of damping is that it not only broadens the amplitude response, but also shifts the amplitude maximum to lower frequency: Supplementary Figure 8.1(a). However, the center of the phase shift always intersect the resonance frequency regardless of damping. Therefore, the phase response of the oscillator provides more accurate information regarding resonance, even in the event that the amplitude response becomes entirely indistinguishable, i.e. overdamped.

In the case that the two oscillators are coupled, we have additional coupling parameters defined as $\omega_{c1} = (K/m_1)^{1/2}$ and $\omega_{c2} = (K/m_2)^{1/2}$. The equation of motion for the coupled system, assuming only the Higgs oscillator is directly driven by the periodic drive, is given by

$$\begin{pmatrix} \omega_1{}^2 + \omega_{c1}{}^2 - \omega^2 + ib_1\omega & -\omega_{c1}{}^2 \\ -\omega_{c2}{}^2 & \omega_2{}^2 + \omega_{c2}{}^2 - \omega^2 + ib_2\omega \end{pmatrix} \times \begin{pmatrix} x_1 \\ x_2 \end{pmatrix} = \begin{pmatrix} \omega_1{}^2 X \\ 0 \end{pmatrix},$$

which gives the solution

$$x_1 = \frac{(\omega_2{}^2 + \omega_{c2}{}^2 - \omega^2 + ib_2\omega)\cdot\omega_1{}^2 X}{\det} \quad \text{and}$$

$$x_2 = \frac{\omega_{c2}{}^2\omega_1{}^2 X}{\det},$$ where det is the determinant of the 2 × 2 matrix above.

While $x_1$ exhibits two resonances arising from the poles in the denominator, it also exhibits an anti-resonance arising from zero in the numerator. An example of the response of the Higgs oscillator in the coupled scenario is shown in Supplementary Figure 8.1(b). Figure 4(a)(b) in the main text is plotted with are $\omega = 0.575$, $\omega_1 = 1 \times \Delta(T)$, $\omega_2 = 0.5 \times \Delta(T)$, $b_1 = 1$, $b_2 = 0.01$, $\omega_{c1} = 1.35$, $\omega_{c2} = 0.5$, where $\Delta(T) = \sqrt{n_s(T)}$ . $n_s$ is the superfluid density in LSCO(OP45) determined from independent London penetration depth measurement. The temperature dependence of $n_s$ for LSCO(OP45) is shown in Supplementary Figure 8.2.

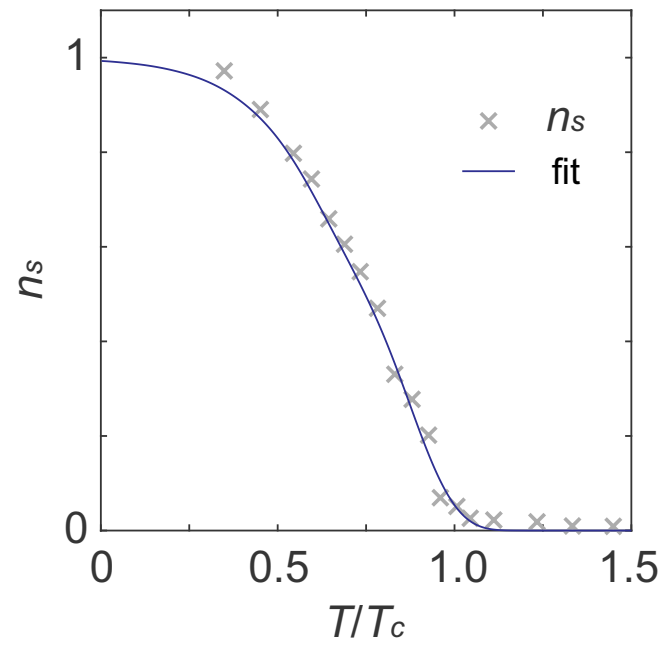

**Supplementary Figure 8.2** Superfluid density in LSCO(OP45). Experimental data (crosses) are obtained from London penetration depth measurement in the terahertz and microwave frequency range. Solid line is a polynomial fit to the experimental data.

## Supplementary Note 9. Extended Anderson pseudospin model

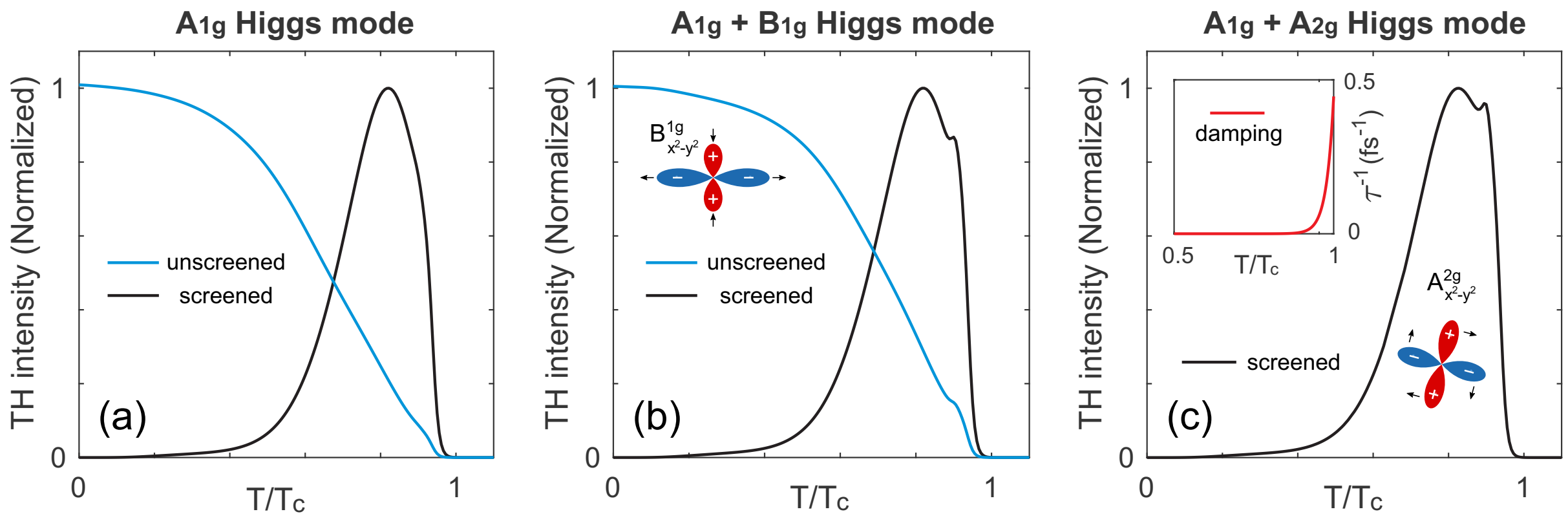

**Supplementary Figure 9** Extended Anderson pseudospin model incorporating an (a) $A_{1g}$, (b) $A_{1g}$ + $B_{1g}$, and (c) $A_{1g}$ + $A_{2g}$ drive of the complex order parameter Δ in the pseudomagnetic field. The inset in (c) shows the temperature-dependent damping function assumed for the model. The drawings in (b)(c) illustrate the $B_{1g}$ and $A_{2g}$ Higgs modes of the *d*-wave order parameter.

One potential candidate for the coupled collective is the non-$A_{1g}$ Higgs mode of the *d*-wave superconducting order parameter. The symmetry of these modes is described as $A_{2g}$, $B_{1g}$, $B_{2g}$, and their motion in momentum space is illustrated in Supplementary Figure 9. More details of these symmetrical modes can be found in (*2, 3*). According to theoretical predictions, the resonance energy of the $A_{2g}$ and $B_{1g}$ Higgs mode is less than the $A_{1g}$ Higgs mode. They might be potential candidate for the new collective mode. THG from the driven Higgs oscillation can be modeled by the Anderson pseudospin model. However, a coupling between the $A_{1g}$ and non-$A_{1g}$ Higgs mode within the Anderson pseudospin model is not straightforward. Therefore, we illustrate below the THG response of an independently driven oscillation of the $A_{2g}$/$B_{1g}$ Higgs mode.

To model the TH response of a periodically driven $A_{2g}$/$B_{1g}$ Higgs mode, we use an extended Anderson pseudospin formalism with a standard BCS Hamiltonian (*4-6*). The Hamiltonian reads $\boldsymbol{H} = \sum_{\mathbf{k}} \boldsymbol{b}_{\mathbf{k}} \boldsymbol{\sigma}_{\mathbf{k}}$, where $\boldsymbol{\sigma}_{\mathbf{k}}$ is Anderson's pseudospin describing the occupation of quasiparticles and Cooper pairs. The pseudomagnetic field reads

$$\boldsymbol{b}_{\mathrm{k}} = \begin{pmatrix} -2\Delta'(t)f_{\mathrm{k}} \\ 2\Delta''(t)f_{\mathrm{k}} \\ 2\epsilon_{\mathrm{k}} \end{pmatrix},$$

with the dispersion $\varepsilon_{\mathrm{k}}$, the energy gap $\Delta(t) = \Delta'(t) + i\Delta''(t) = \frac{W}{N}\sum_{\mathrm{k}} f_{\mathrm{k}} \left(\langle\sigma_{\mathrm{k}}^{\mathrm{x}}\rangle(t) - i\langle\sigma_{\mathrm{k}}^{\mathrm{y}}\rangle(t)\right)$, and the gap symmetry function $f_{\mathrm{k}}$. The time evolution is described by Bloch equation $\dot{\sigma}(t) = \boldsymbol{b}_{\mathrm{k}} \times \sigma_{\mathrm{k}}(t)$.

The coupling to the electromagnetic field $\mathbf{A}(t) = \mathbf{A}_0 \sin(\omega t)$ is considered by minimal substitution $\epsilon_{\mathrm{k}} \to \epsilon_{\mathrm{k}-e\mathbf{A}(t)}$. To model a driving of the $A_{2g}$ ($B_{1g}$) Higgs mode, we use a phenomenological approach, where the gap symmetry is modulated periodically in a different symmetry channel however with the same time-dependence as the usual driving with light. Such an asymmetric driving may result experimentally from a small in-plane component of the driving light wave vector or higher order couplings. To this end, we replace the symmetry function $f_{\mathrm{k}}$ in the pseudomagnetic field $\boldsymbol{b}_{\mathrm{k}}$ by an effective time-dependent symmetry function $f_{\mathrm{k}}^{\mathrm{Q}}(t) = f_{\mathrm{k}} + \delta|A(t)|^2 f_{\mathrm{k}}^{\mathrm{Q}}$, where $f_{\mathrm{k}}^{\mathrm{Q}}$ has a different symmetry from $f_{\mathrm{k}}$. Finally, we include the screening by a temperature-dependent driving amplitude $\mathbf{A}_0(T)$ (given by $I_{\mathrm{FH}}(T)^{1/2}$) and a temperature-dependent damping modeled by a relaxation time $\tau(T)$ in the Bloch equations (Supplementary Figure 8(c) inset). The assumption of a temperature-dependent damping follows from experimental observation of a diverging pair-breaking rate as $T$ approaches $T_c$ (*7, 8*).

Finally, the Bloch equation reads

$$\dot{\sigma}(t) = \begin{pmatrix} -2\Delta'(t)f_{\mathrm{k}}^{\mathrm{Q}}(t) \\ 2\Delta''(t)f_{\mathrm{k}}^{\mathrm{Q}}(t) \\ \epsilon_{\mathrm{k}-e\mathbf{A}(t)} + \epsilon_{\mathrm{k}+e\mathbf{A}(t)} \end{pmatrix} \times \begin{pmatrix} \sigma_{\mathrm{k}}^{\mathrm{x}}(t) \\ \sigma_{\mathrm{k}}^{\mathrm{y}}(t) \\ \sigma_{\mathrm{k}}^{\mathrm{z}}(t) \end{pmatrix} - \frac{\sigma_{\mathrm{k}}^{\mathrm{z}}(t) - \sigma_{\mathrm{k}}^{\mathrm{z}}(0)}{\tau(T)}.$$

We numerically solve the Bloch equation self-consistently together with the gap equation. To obtain the THG intensity in the same polarization direction as the electromagnetic field, we evaluate $I^{\mathrm{TH}} \propto |j^{(3)}(3\omega)|^2$, where the nonlinear current is $j^{(3)}(t) \propto \Delta_0 A(t)\delta\Delta(t)$.

For the results shown in Supplementary Figure 9, we use $\boldsymbol{\delta} = 0.04$ for the $A_{1g} + B_{1g}$ Higgs mode case and $\boldsymbol{\delta} = 0.08$ for the $A_{1g} + A_{2g}$ Higgs mode case. The superfluid density of LSCO(OP45) is used for $\Delta$(T).

Supplementary References:

1. Kovalev, S. *et al.* Probing ultra-fast processes with high dynamic range at 4th-generation light sources: Arrival time and intensity binning at unprecedented repetition rates. *Structural Dynamics* **4**, 024301 (2017).

2. Barlas, Y. & Varma, C. M. Amplitude or Higgs modes in d-wave superconductors. *Phys. Rev. B* **87**, 054503 (2013).

3. Schwarz, L. *et al.* Classification and characterization of nonequilibrium Higgs modes in unconventional superconductors. *Nat. Comm.* **11**, 287 (2020).

4. Anderson, P. W. Random-Phase Approximation in the Theory of Superconductivity. *Phys. Rev.* **112**, 1900–1916 (1958).

5. Matsunaga, R. *et al.* Light-induced collective pseudospin precession resonating with Higgs mode in a superconductor. *Science* **345**, 1145 (2014).

6. Tsuji, N. & Aoki, H. Theory of Anderson pseudospin resonance with Higgs mode in superconductors. *Phys. Rev. B* **92**, 064508 (2015).

7. Parham, S. *et al.* Ultrafast Gap Dynamics and Electronic Interactions in a Photoexcited Cuprate Superconductor. *Phys. Rev. X* **7**, 041013 (2017).

8. Reber, T. J. *et al.* Pairing, pair-breaking, and their roles in setting the $T_c$ of cuprate high temperature superconductors. https://arxiv.org/abs/1508.06252 (2015).